# Multi-Assay Antibody Developability Prediction with Shared Structure-Language Representations

Daria B. Kokh*

## Abstract

Early assessment of antibody developability is critical in drug discovery, but computational approaches remain limited by the scarcity of experimental data. We explore two strategies for working within that limit.

First, we designed per-residue physicochemical descriptors encoding the local structural environment, including new terms such as directional charge asymmetry and screened electrostatic potential. These feed a convolutional network, combined with protein language model (PLM) embeddings in a two-stream architecture: the two are encoded and predicted independently and merged only at the prediction stage. Second, a shared trunk with per-assay heads passes information between related assays.

Both are implemented in the Multi-Assay Developability Model (MADM) and evaluated on eight assays: HIC, heparin affinity (HAC), AC-SINS, and polyspecificity ELISA against BVP, dsDNA, insulin, ovalbumin and soluble membrane proteins (SMP), with labels from 29 open sources.

Structural descriptors match or exceed PLM embeddings (AbLang2, ESM2) on several assays; combining the streams gives a small but consistent gain in accuracy and reduces seed-to-seed variance. Multi-assay training raises Spearman rho for ovalbumin and SMP by about 0.1, where assays correlate and share sequence space.

Under cluster-split cross-validation the strongest results are for HAC, ovalbumin, insulin, SMP and HIC (rho = 0.70-0.82); AC-SINS, BVP and dsDNA are weaker (0.50-0.55). In these three, sequence clusters largely coincide with data source, so a held-out fold measures transfer to differently calibrated measurements as well as to new sequences. On the 80-antibody GDPa3 hold-out, performance falls by rho around 0.1 for all but SMP (0.22). Transfer to the VHH panel of Ginkgo GDPa5 gives comparable results, and scFv validation against binary SMP labels yields F1 = 0.89, AUC = 0.95.

* Email: db.kokh@gmail.com

# 1. Introduction

Screening early-stage antibody candidates for developability is an important step toward selecting molecules that can successfully progress further development, covering both manufacturability and therapeutic performance. However, since experimental assays are often low-throughput, costly, and time-consuming, they are not routinely applied at the early stages of antibody discovery, where large candidate libraries need to be evaluated. This is especially true for the de-novo or computationally designed process required screening large number of candidates that may have low similarity, and therefore more diverse physicochemical properties relative to well-characterized antibodies.

One of the most common computational directions is to quantify a set of physicochemical characteristics of molecules associated either in general terms, such as Therapeutic Antibody Profiler (TAP [1]) or for specific liabilities, as in the spatial aggregation propensity and spatial charge map family (SAP/SCM)[2], [3] , which provide rough general assessment for each molecule suitable for go or no-go conclusion.

A more focused assessment targets individual properties by coupling descriptors to measured data from a specific developability assay, such as HIC for hydrophobicity. The typical workflow computes a large panel of sequence- and structure-based descriptors, often generated with commercial software packages, and then reduces their dimensionality separately for each assay. Several assay-specific methods of this kind have performed well: HIC-derived propensity scale[4] ; MD-derived molecular surface descriptors benchmarked against viscosity, aggregation, HIC, clearance, heparin retention, and polyspecificity,[5][6], and their larger-scale counterpart PROPERMAB[7]. Across these studies, hydrophobic surface patches and charge properties (such as pI) are generally the highest-weighted features[8]. Each method nevertheless uses a different descriptor set, and performance typically degrades on antibodies outside the training set and on measurements made under a different experimental setup. Model generalizability is a common limitation: SCM classifies IgG4 viscosity with ~80% accuracy on a 14-antibody set, but shows almost no correlation on a substantially larger panel[7], good cross-validation performance for many assays drops by Spearman ~0.3 as applied to unseen data set[9]

Trained on hundreds of millions to billions of protein or antibody sequences, PLMs capture information related to general sequence and physicochemical patterns. Although a recent benchmark showed that unsupervised, zero-shot, developability assay prediction based on PLM embedding is not efficient [10], good performance was reported for supervised models based on either embedding or including fine-tuning and learn from particular assays, which showed promising results for predicting specific assay outcome [11],[12],[13],[14],[15] with domain-adaptive fine-tuning working better than a general-purpose model[11]. Often structure-based descriptors are used as additional supportive parameters for PLM fine-tuning, and PLM-based models can outperform structure-based ones [16]. On the other hand, the FLAb2 [10] evaluation of 30 AI and biophysical models on over 4M antibodies across 32 studies concludes that what a PLM contributes on these tasks is a measure of how unusual a sequence is relative to the repertoire it was trained on, rather than knowledge of the biophysics, consistent with another observation that failing sequences carry systematically higher PLM perplexity[11].

Supervised models are inseparable from the generalizability problem, because they require assay-specific datasets large enough to support it. Developed on datasets of very different type and size, frequently proprietary, models usually are not directly comparable. One of the first systematic efforts to establish benchmark datasets was reported by Jain et al. [17], and this work was more recently extended by Ginkgo Bioworks. [18] The latter includes five developability assays used to define a blinded community benchmark with a public training set of GDPa1[18] and an 80-antibody held-out test set[9]. That benchmark shows that predictive performance remains limited for most of developability assays, with best held-out ρ=0.708 for hydrophobicity HIC and around 0.31-0.39 for tasks such as AC-SINS, SMP (PR-CHO), and Tm2 (DSF). More revealing than the absolute values is the collapse between cross-validation and held-out performance for the interaction-dependent properties by Δρ = 0.2-0.3, which clearly show the generalizability problem of models.

In the present study we explore three questions that may help address the generalizability problem while still working with a limited dataset: (i) whether representing the spatial arrangement of physicochemical patterns on the antibody structure at per-residue resolution improves assay prediction; and (ii) whether assays with a related biophysical basis can pool complementary information that improves prediction for each individual assay, (iii) whether data combined from different sources can be used for training and what is the limitations of such datasets.

To this end we (1) assembled developability measurements from 29 independent sources into a single training corpus, introducing normalisation strategies for combining them, and quantified the limitations; (2) built a set of per-residue structure-based features, including commonly used descriptors such as per-residue charge, extended them with additional designed features;  instead of the commonly used reduction them to molecule-level descriptors, we retained spatial biophysical properties per residue on a common IMGT frame and used them to train a small convolutional network combined with a protein language model by late fusion, so that the structural and sequence representations are encoded independently and merged only at the prediction stage; (3) showed that this structural stream carries information comparable to the language model, exceeding it on some assays, and that combining the two independent streams improves accuracy and reduces variance across seeds; (4) built a multi-assay model that shares both physics-based and sequence-based encoders across assays while keeping per-assay prediction heads, and established the conditions under which the resulting gain does not persist; (5) assessed cluster-based cross-validation as a model-selection protocol and identified where it provides a stricter test and where it instead measures inter-source calibration; and (6) evaluated the resulting models on four external hold-out sets, including transfer to the VHH and scFv formats.

## 2. Related Work

### 2.1 Physicochemical descriptors.

A range of approaches incorporate physicochemical information into antibody developability assessment by computing properties at the residue or molecular-surface level using predefined physicochemical rules. They generally account for the local environment by aggregating contributions from neighbouring residues, either along the sequence or within a defined region of the three-dimensional structure (e.g. only CDR loops, or only VH/VL chain). The resulting local profiles are then typically summarized into molecular-level scores. The most widely used are TAP [1] -  a set of descriptors for empirical developability guidelines; SAP- local aggregation risk identifying hydrophobic surface patches that may promote aggregation[3]; SCM- spatial clustering of exposed atomic charges that was developed particularly for predicting high-concentration antibody viscosity[2]; Aggrescan3D[19] to identify structural aggregation-prone regions; descriptors generated by commercial software, such as MOE; MolDesk that based on the hydrophobicity and electrostatic field mapped of the molecular surface and has been benchmarked against viscosity, aggregation, HIC, clearance, heparin retention[5]. Such approaches provide physically interpretable descriptors, but the final aggregation step inevitably compresses some of the residue-level and spatial information.

The present work uses several descriptors designed following the approaches of TAP[1], SAP, SCM [2],[3], as well as the residue-level field mapping of MolDesk[5], alongside widely employed general features, such as pI, total and per-fragment charge and dipole moments, secondary structure, and light-chain type, adding also some new or derived characteristics (directional charge asymmetry, hydrophobic patch geometry, screened electrostatic potential and some others, see Table 2).

## 2.2 Combining PLM embedding with physicochemical descriptors.

Two routes dominate the published work that connects such descriptors to measured assay data: either case-study of per-residue surface aggregation or hydrophobicity propensity effect, or molecule-level aggregated features fed into a machine-learning model. Few studies retain spatial resolution inside the model: PfAbNet feeds the electrostatic potential surface of the variable region into a convolutional network for prediction of high-concentration viscosity. [20] We likewise use descriptors as input features to a convolutional network, but retain multiple physicochemical fields per residue rather than a single channel.

Hybrid designs combining embeddings with molecule-level descriptors are often employed to inject structural information necessary for developability prediction. One of the examples is combining AbLang2[21] and ESM2[22] with MD-based descriptors, DeepSP, and the isoelectric point and fed to tabular neural networks, showing that IgG-trained models transfer zero-shot to VHH-Fc for surface-driven properties (HIC $\rho$ = 0.63, AC-SINS $\rho$ = 0.62). [23] Fusion of MOE physicochemical descriptors plus ESM-2 and AbLang embeddings with the fused model outperforms a purely structural CNN baseline, [12] although the gain from adding a language representation to good structural descriptors is modest ($\Delta\rho<0.1$). The same pattern holds for non-specific binding, where embeddings and interpretable sequence-derived biophysical descriptors reach comparable accuracy and are most reliable in combination.[24] A multimodal framework combines structure, sequence and composition descriptors across four datasets covering conformational stability, colloidal stability, viscosity and solubility.[25] Unlike these designs, which concatenate descriptors and embeddings into a single representation, we keep the two streams separate and merge them only at the prediction stage, with a fitted per-assay weight. The practical reason is their difference in dimensions: a frozen language-model embedding supplies dimension of several hundreds to thousand channels per residue against 66 designed channels, so in a concatenated representation a shared head can satisfy most of the training loss from the larger block alone, and the structural contribution is largely missing. Late fusion requires each stream to produce its own prediction, so neither can be carried by the other, and the fitted weight states directly how much each contributes to a given assay.

A larger data set also permits building a model that learns the spatial distribution of physicochemical properties from amino-acid type and position on a graph built from the predicted structure. Applied to size-exclusion chromatography, this model was shown to perform comparably to a protein language model on approximately 1200 labels, which is close to the practical minimum given the parameter count of a typical graph network [26]. Combining the graph network with the language model left mean performance unchanged while increasing variance across folds. Another related architecture couples a fine-tuned language model with residue-aligned AlphaFold context encoded as residue graphs and fused by multi-stream attention, for clone self-interaction measured by biolayer interferometry for 1499 IgGs and 841 VHHs sequences, however the result just minimally outperformed an interpretable descriptor model on the same structures. [27]

## 2.3 Combining labels from multiple assays.

Finally, the use of alternative assay labels to improve prediction has recently been explored by Moller et al. [23] , who added measured HIC and heparin (HAC) values as additional input features to a polyreactivity (PR-CHO) model, raising ρ by about 0.1. Because the auxiliary labels are inputs, their values must be available for every molecule scored, which defines a tiered screening cascade in which the cheaper surface assays are measured first. It is worth noting that this route is validated within a single platform: all measurements in that study, training and test alike, were generated on one standardised system[18], so a value supplied at inference is already on the scale the model was trained on, and the model may learn platform-specific assay-to-assay relationships that need not hold when the auxiliary measurement comes from elsewhere. In the present work the auxiliary assays instead enter as additional training tasks sharing the encoders with the target, so their labels are required only during training and prediction needs nothing beyond sequence and structure.

Jiang et al. [28] likewise use a shared backbone with per-assay heads, trained jointly on eight measured GDPa1 assays. Our architecture differs in that training is asymmetric: a separate model is fitted for each target assay, with the remaining assays entering as weighted auxiliary tasks rather than as co-equal objectives. This allows the auxiliary set and its weighting to be chosen per target.

## 3. Data

Data were collected from 29 publicly-available sources (Table S1) including GitHub, articles, and patents. Each antibody was represented by its paired heavy- and light-chain variable sequences (VH, VL) and annotated with experimental measurement: AC-SINS, Hydrophobic Interaction Chromatography (HIC), Heparin Affinity Chromatography (HAC), polyreactivity reagent binding: BVP, dsDNA, insulin, ovalbumin, and soluble membrane proteins (SMP/PR-CHO) ELISA.

As measurements were pooled from several independent sources, they may differ in two respects: (i) the experimental conditions under which the assay was run and metrics used to generate reported labels, and (ii) the antibody format (e.g. Fc subtype). We did not address the second issue because of too few labels with the same Fv and different Fc fragments to estimate reliably a format effect. For the first one, the labels were rescaled to a common baseline per assay using anchor-antibodies measured in more than one source (script for data rescaling is provided with source codes). Where no anchor was available, the original values were retained, so discrepancy between source is still possible. For HIC we additionally used rescaling based on experimental conditions: measurements with flow-rate 0.75 ml/min or less were scaled to the 1ml/min using common therapeutic antibodies from Jain et al. [17] and Estep et al. [29].

For each assay an assay-specific threshold was defined that separates "bad" from "good" antibodies largely following recommendation of Jain et al. [17]. It was used only to derive a binary label for classification-style evaluation.

Data characteristics for each assay are summarised in Table 1. Three aspects of the data structure are important for the result interpretation. First, the number of contributing sources varies from just 2 (HAC) to 19 (AC-SINS), and sequence vary by similarity. Our analysis showed that in several assays clusters built by sequence similarity (see Sec. Method below) are strongly associated with source so that a cluster-based split becomes actually in some cases a source-based split. Therefore, under cluster CV the model may learn across sequences as well as across laboratories, and conversely, it may exploit source-specific offsets rather than biophysical properties of molecules. We quantify this with Normalised Mutual Information between the cluster and source labelling of the same molecules (NMI = 0: independent; NMI = 1: identical partitions) shown in Table 1 for each assay. The most correlated assay-sequence data sets are BVP, dsDNA, followed by Insulin and AC-SINS.

Second, the number of clusters and the maximum cluster size vary widely across assays. HIC, SMP and Ovalbumin are dominated by singletons, whereas dsDNA, AC-SINS, and BVP comprise a few large clusters and a correspondingly high probability *P* that two randomly drawn molecules share a cluster. As will be discussed later, this causes the gap between random-split and cluster-based performance.

Third, labels for antibodies measured in more than one source allow the assay-limited prediction accuracy to be estimated. The least accurate is AC-SINS assay, where the n-weighted rank agreement between sources is 0.68, placing a ceiling on any model trained on it. The apparent measurement noise can be estimated per assay (Table 1) and accounted for during training through the noise floor.

# 4. Methods

## 4.1 Data preprocessing

3D Fv structures were generated with ABodyBuilder2 (NanoBodyBuilder2 for VHH)[30], and used to compute per-residue biophysical features together with whole-molecule global descriptors. For each antibody the per-residue feature matrix was aligned to a common IMGT frame[31], stored with the model with absent positions zero-filled and excluded by the residue mask.

Target labels were transformed before training. We found that the best performance was for log1p transform followed by z-scoring. The transform parameters (mean, standard deviation, and the log flag) were fit on the training labels of each split only and applied to the corresponding validation labels, so that no information from the held-out set entered the transformation. All reported metrics were computed after inverting the transform back to the original assay units.

**Table 1 Datasets and sequence-cluster structure.** Good/bad thresholds follow Ref.[17] where available. Noise threshold is the assay-specific measurement floor, below which values are treated as unreliable. Sequence clusters are the connected components of the graph linking two antibodies whose VH or VL exceed 0.8 4-mer Jaccard similarity; all members of a cluster are assigned to the same cross-validation fold. P is the probability that two randomly drawn molecules fall in the same cluster (P = 0- unconstrained random split; larger P means a stricter, coarser partition). Singletons are molecules alone in their cluster and are therefore unconstrained. NMI, Normalised Mutual Information between cluster and source (NMI =0: independent, NMI =1: cluster and source are the same partition)

| Assay | Threshold good/ bad | Assay noise threshold | Data Set size | # of clus-ters | max cluster size | P | molecules in singletons, % | # of Sources | NMI (cluster, source) | assay-limited predictive accuracy * |
|---|---|---|---|---|---|---|---|---|---|---|
| HIC | 11 min | - | 771 | 516 | 66 | 0.017 | 63.3 | 10 | 0.405 | 0.98 |
| BVP | 4** | - | 346 | 140 | 95 | **0.081** | 31.4 | 10 | **0.657** | - |
| AC-SINS | 11 nm | 3 | 620 | 248 | 109 | **0.047** | 37.7 | 19 | **0.663** | 0.68 |
| Insulin | 0.5** | 0.1 | 297 | 152 | 31 | **0.032** | 45.4 | 9 | **0.582** | - |
| Ovalbumin | 0.5** | - | 323 | 278 | 15 | 0.003 | 81.4 | 3 | 0.250 | 0.78 |
| dsDNA | 0.5** | - | 159 | 42 | 31 | **0.084** | 18.8 | 7 | **0.788** | - |
| SMP (PR-CHO) | 0.3** | 0.1 | 382 | 329 | 25 | 0.005 | 82.9 | 4 | 0.270 | 0.72 |
| Hep (HAC) | 4** | 0.05 | 171 | 159 | 2 | 0.002 | 90.1 | 3 | 0.222 | - |
| GDPa1[18] | - | | 246 | 222 | 9 | 0.002 | 87.7 | 1 | - | - |
| GDPa3[9] | - | | 80 | 80 | 1 | 0.000 | 100 | 1 | - | - |

*****The assay reproducibility cap computed from antibodies measured independently in more than one source. Spearman ρ between their values on the overlapping set; **score

## 4.2 Multi-Assay Model

The general architecture is shown in Fig. 1 and comprises two streams: a masked convolutional trunk for the structural features (Fig. 1B) and a projection block for the AbLang2 (or ESM2) embeddings (Fig.1C), giving two complementary per-residue representations combined by late fusion.

### 4.2.1 Structural stream.

31 structure-derived per-residue biophysical descriptors (66 channels in total) are computed from the modelled Fv or VHH structures (see Table 2) and assembled into an [N, L, F] tensor, where N is the number of molecules, L = 290 the number of IMGT frame positions and F the number of channels, together with a per-residue mask. Each channel was standardised to zero mean and unit variance using statistics estimated on the training partition and applied unchanged to the validation partition; padded positions were zeroed by the mask. Seven global descriptors were appended, describing VH and VL charge, isoelectric point and light-chain type (Table 2).

Three 1-D convolutional blocks (Fig. 1B) encode the residue axis, followed by masked attention pooling to a fixed-length vector, which is concatenated with the global descriptors and mapped to a scalar by an MLP head.

### 4.2.2 Language-model stream.

A sequence-only model built on frozen per-residue embeddings: AbLang2-paired (480-dimensional) or ESM2-650M (1280-dimensional). For ESM2 each chain was encoded separately and the per-residue embeddings concatenated along the sequence axis. These pass through a linear, MLP or convolutional block and are pooled by attention or mean/max to a molecule-level representation, then mapped to the assay value by an MLP head. Target transform, cross-validation and evaluation are identical to the structural stream, so the two encoders are directly comparable. Optimisation is described in Sec. 4.2.5 and is identical for both streams.

### 4.2.3 Two streams.

The two streams are combined by late fusion. Each encoder is pooled independently to a fixed-length vector; the structural vector is concatenated with the global descriptors, which the sequence stream does not see. Each stream has its own per-assay MLP head producing a scalar prediction, and the two are combined with a learned per-assay weight, $p = w \cdot p_feat + (1- w) \cdot p_ab$ with $w = \sigma(w_logit)$, so that each assay can rely on the two streams to a different degree. The representations themselves never mix.

### 4.2.4 Multi-task model.

The multi-task model shares both encoders across assays and attaches a small per-assay regression head, so that one common molecule-level representation is mapped to each assay's value. Both encoders are warm-started from the corresponding single-stream models fitted on the same fold; warm starting sets the initial weights only, and encoders, heads and fusion weights are then trained jointly on the target together with a set of auxiliary assays.

For each task, only molecules carrying a label for that task contribute to its loss. A molecule with an auxiliary measurement but no target value therefore still trains the shared encoders through its auxiliary head, though not the target head, which receives gradient only from target-labelled molecules.

The target's weight in the joint loss is the number of auxiliary assays, each auxiliary carrying weight 1, so that the target accounts for half the total loss at initialisation. The auxiliary weights are then learned as homoscedastic uncertainties[32] while the target's weight remains fixed, allowing the model to down-weight auxiliaries that transfer poorly; the target's share of the loss consequently increases during training and the model degrades gracefully toward single-task behaviour.

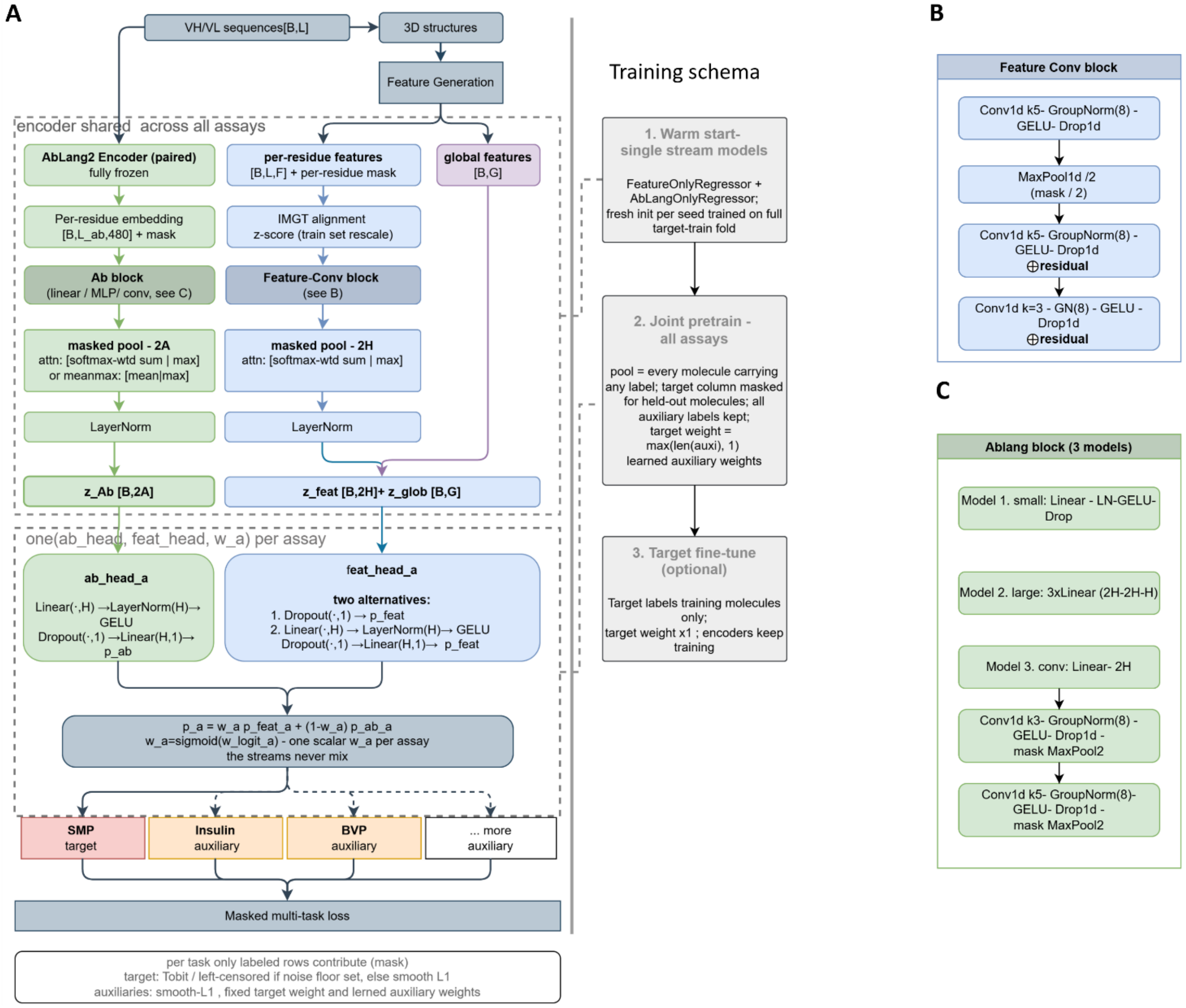

**Fig. 1** **(A)** VH/VL sequences yield a predicted structure, from which per-residue biophysical features (IMGT-aligned, z-scored on training data) and global descriptors are computed, and a frozen AbLang2 paired embedding. Two encoders, shared across all assays, reduce each to a fixed-length vector; global features enter the structural stream only. Each assay has its own pair of heads and a single learned blend weight, and predictions are combined by late fusion, $p_a = w_a \cdot p_feat_a + (1 - w_a) \cdot p_ab_a$, $w_a = \sigma(w_logit_a)$; the streams are never concatenated. Training minimises a masked multi-task loss over labelled rows only, in three stages: single-stream warm start, joint pre-training on all assays with the target column masked for held-out molecules, and a target-only fine-tune (optional). **(B)** Structural encoder: k = 5 convolution, mask-aware max-pooling by two, then k = 5 and k = 3 residual convolutions, all with GroupNorm (8), GELU and 1-D dropout. **(C)** The three AbLang encoder variants: linear projection, three-layer perceptron, or projection followed by two convolutional blocks with mask-aware pooling.

### 4.2.5 Training.

**Each stream** was first fitted alone as a single-task model for 200 epochs, and the two-stream model initialised from that pair so that it began at the plain average of the two streams. All fits used Adam at a learning rate of $1 \times 10^{-3}$ with weight decay $5 \times 10^{-3}$, squared-error loss, gradients clipped to a maximum norm of 1.0, and a cosine-annealing schedule from the base rate to $1 \times 10^{-5}$ over the epoch budget.

**The fused model** was trained for at most 260 epochs, stopping after 20 epochs without improvement on the validation fold. Epoch selection used the target's validation loss, computed on the target term alone. The weights of the best-scoring epoch were restored before evaluation.

Where a noise floor was defined for an assay, predictions and measurements were clipped at it before the per-epoch rank metrics were computed, so that ordering below the detection limit is not reported as signal. The reported metrics are the uncensored ones in original assay units, but the censored pair is computed and stored alongside them.

For each split, five models were trained with different random seeds defining weight initialisation and data ordering, with partitions held fixed. Seed predictions were averaged into per antibody and metrics computed on that ensemble.

For the final models the epoch budget was fixed to the median best epoch observed across CV runs, separately for each warm-started stream, and five models were then fitted on all labelled molecules with no validation fold. Early stopping is not available there.

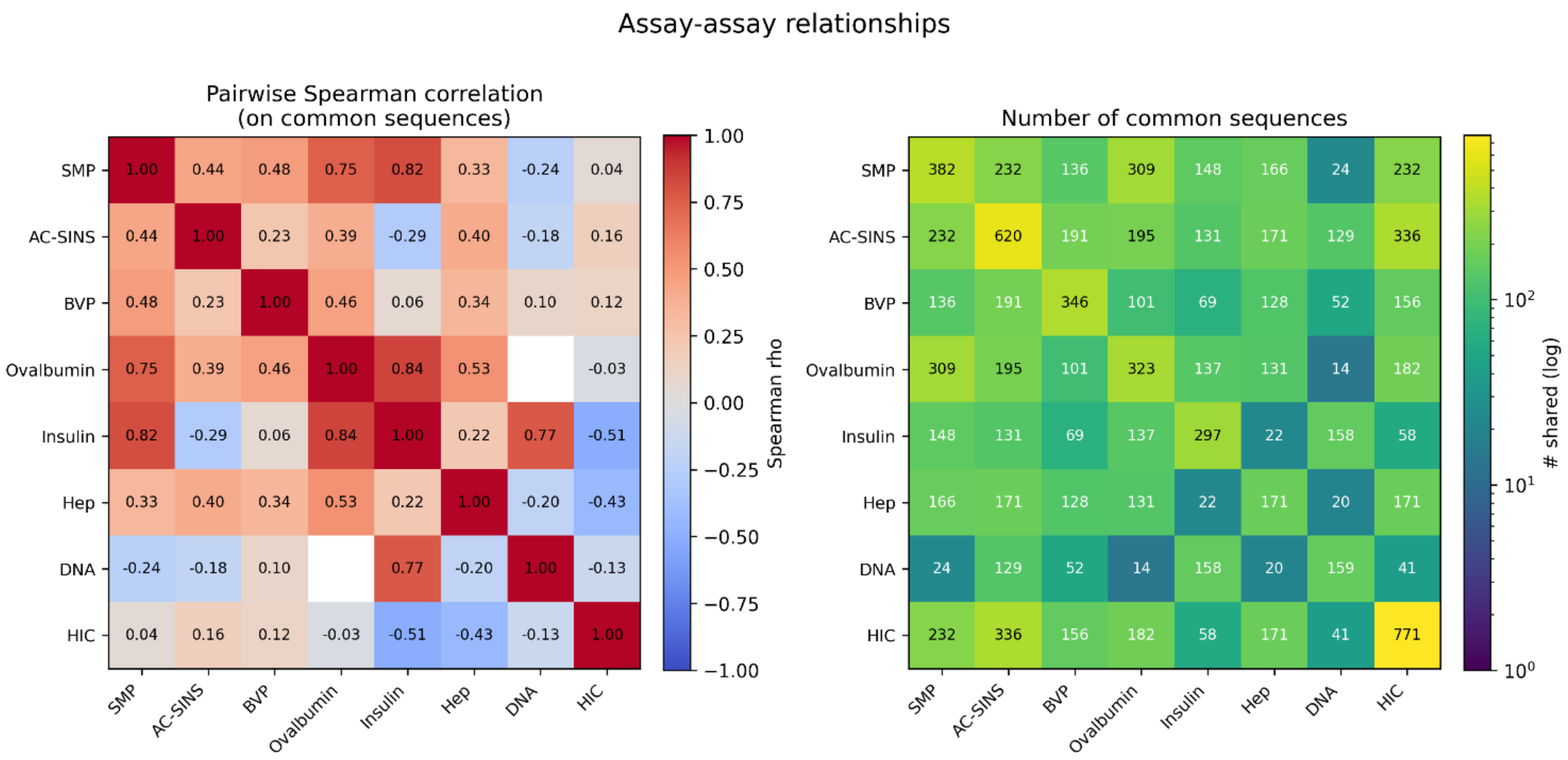

**Fig.2** Pairwise relationships between developability assays. Assay–assay agreement and data overlap across the eight assays (SMP, AC-SINS, BVP, Ovalbumin, Insulin, Hep, DNA, HIC). A: Pairwise Spearman ρ on common VH/VL sequences. For each assay pair, Spearman's ρ was computed on the subset of antibodies measured in both assays. Correlations are shown only for pairs with sufficient overlap (n ≥ 10 co-measured antibodies). Most assays are weakly-to-moderately positively correlated, reflecting shared biophysical drivers of non-specificity, with a few strong pairs. B: Number of shared sequences. Each cell gives the count of antibodies measured in both assays (log colour scale); the diagonal gives the total number measured in that assay.

**Table 2** Per-residue features; hydrophobicity score was used from ΔG (water—lipid bilayer)[33] or from experimental HIC retention [4]

| # | Type | Descriptor | Dimension/ Thresholds(Å) | Definition |
|---|---|---|---|---|
| 1 | local | angles | [L,3] | Backbone torsion angles. |
| 2 | local | sasa | [L] | Solvent-accessible surface area (SASA). |
| 3 | Local | sasa_sch | [L] | Solvent-accessible surface area (SASA) of side-chains |
| 4 | local | neighbors | [L,6] | Structural neighborhood: distance to each of the nearest residue j, defined as the minimum distance from the Cα of residue i to any heavy atom of residue j; residue i itself and its immediate sequence neighbors are excluded. |
| 5 | local | hydro_sum | [L,2]/[6,12] | Σ(hy_score_j × exposure_sch_j) over residues within a distance threshold, where exposure_sch = sasa_sch / sasa_ref is the side-chain exposure relative to a fully exposed residue. |
| 6 | local | charge_sum | [L,2]/[6,12] | Σ(q_i·q_j / d²) over exposed residues only, where q is charge and d the inter-residue distance; +1- Lys/Arg, +0.1-His,-1- Asp/Glu. |
| 7 | local | h_bonds | [L] | Number of hydrogen bonds. |
| 8 | local | contacts | [L,3] | Number of per-residue contacts within a distance threshold, using all heavy atoms. |
| 9 | local | contacts_interchain | [L,2]/[6,12] | Number of per-residue interchain contacts within a distance threshold, using all heavy atoms. |
| 10 | local | aromatic_sum | [L,2]/[6,12] | Σ(sasa_sch_j) over aromatic residues within a distance threshold of residue i, where sasa_sch is side-chain exposure |
| 11 | local | q_dipole | [L,4]/[6,9,12,16] | Quasi-dipole = \|q_A- q_B\| × d, where d = \|\|centroid_A − centroid_B\|\| (3D Euclidean) and q_A, q_B are the summed charges of two patches. A patch = surface residues within the threshold distance of CA_A/ CA_B; the patch anchors A and B are the surface residues whose COM→Cα direction is most opposite along the axis through the COMs |
| 12 | local | potential_sites | [L,3] | Number of possible contacts × exposure of the contact (H-bond donor, acceptor, polar, and charged atoms included). |
| 13 | local | cdr | [L,6] | Membership in each of the CDR loops (one indicator per loop). |
| 14 | Local | charge_cdrs | [L] | Charge of CDR, assigned to each residue belonging to that loop |
| 15 | Local | ss_num | [L] | Secondary structure (loop / helix / β-strand). |
| 16 | Local | cdr_fw | [L] | Number of CDR–framework bonds (H-bonds, polar bonds) |
| 17 | Local | q_dipole_h_min | [L,3]/[6,9,12] | Hydrophobic dipole = min(\|h_A\|, \|h_B\|) × d, where h = hydrophobicity score × side-chain SASA; d as in q_dipole. |
| 18 | Local | q_dipole_h_abs_diff | [L,3]/[6,9,12] | Hydrophobic dipole = \|h_A- h_B\| × d; h and d as above |
| 19 | Local | q_dipole_h_product | [L,3]/[6,9,12] | Hydrophobic dipole = \|h_A · h_B\| × d; h and d as above |
| 20 | Local | q_dipole_h_sum | [L,3]/[6,9,12] | Hydrophobic dipole = \|h_A + h_B\| × d; h and d as above |
| 21 | Local | flatness_score_hy | [L] | Flatness/hydrophobicity score: high- residue sits on a flat hydrophobic surface; low- residue is protruding, buried, or hydrophilic. |
| 22 | Local | flatness_score | [L] | Flatness score: high = residue on a flat surface; low = protruding, buried, or hydrophilic. |
| 23 | Local | hydro_patch_sizes | [L] | Size of the hydrophobic surface patch the residue belongs to. Exposed hydrophobic residues (hydrophobicity × SASA > 0.25, side-chain exposure > 0.1) are linked when their Cα atoms lie within 10 Å, and patches are the connected components of that graph. Each residue is assigned the number of residues in its patch; residues in no patch, or forming an isolated single-residue patch, score 0. |
| 24 | Local | salt-bridge_bonded_score | [L] | Max over charged atoms that are in salt bridges of (100 if fully buried, else SASA); 0 if the residue is uncharged or has no salt-bridged atoms |
| 25 | Local | salt-bridge_unused_score | [L] | As salt_bridge_bonded_score but over charged atoms not in salt bridges; 0 if uncharged or all charged atoms are salt-bridged. |
| 26 | Local | pos_potential | [L,2]/[4,16] | Positive/negative electrostatic potential for residue i, evaluated at several probe points (o_i) displaced outward along the side-chain-tip direction: V(p_i) = Σ_j (q_j·w_j)/(ε_w·d_ij) · exp(-d_ij/λ_D), where d_ij = distance from probe i to source j's charge centroid, w_j = exposure of source side-chain residue j, ε_w = water dielectric (~80), λ_D- Debye screening length (~10 Å at physiological salt). |
| 27 | Local | neg_potential | [L,2]/[4,16] | |
| 28 | local | hydro_full_cdr | [L] | Positive hydrophobicity score from the Jain CDRs only |
| 29 | Local | charge_exp | [L] | Residue charge multiplied by SASA, if SASA > 25% of the residue's free (fully exposed) SASA |
| 30 | Local | charged_patch_area_pos | [L,2]/[6,12] | Total exposed area of pos./neg. charged residues within a distance threshold. |
| 31 | Local | charged_patch_area_neg | [L,2]/[6,12] | |
| 32 | global | vh_charge,vl_charge | [2] | Total charge of the chain |
| 33 | global | Ionization, | [2] | Total charge at pH 7.5 , pH 6; PROPKA [34] |
| 34 | global | pI_folded/ pI_unfolded | [2] | pI (PROPKA [34]) |
| 35 | global | is lambda | [1] | Light-chain type; using ANARCI [35]) |

**We deliberately avoided exhaustive per-model hyperparameter optimisation**, varying only a small number of hyperparameters by hand and guided mainly by the size of the available data. For the assays with roughly 300 or more measurements (AC-SINS, HIC, BVP and SMP) we widened the hidden layer of the feature stream to 32–48 units, against 16 elsewhere; for HIC, with 771 measurements, we additionally enabled the convolutional/attention block in the sequence-based stream (the complete list of hyperparameters is in Table S2). Performance depends strongly on the training set and validations sets, so a configuration tuned exhaustively cannot be expected to be transferable.

**Auxiliary assays** for each target were selected on two criteria: the target–auxiliary Spearman correlation on co-measured molecules (Fig. 2A) and the number of sequences measured in both (Fig. 2B). The largest overlap is between SMP and Ovalbumin, which share many identical and highly similar sequences (Fig. 3) and are also the most strongly correlated pair (ρ = 0.75); AC-SINS correlates weaklier with SMP and with BVP (ρ = 0.44 and 0.23). For the five-fold cross-validation we deliberately applied a lower threshold, admitting every auxiliary correlating with the target above ρ = 0.3 (Table S3), in order to test where the benefit breaks down rather than to tune the auxiliary set per assay. Several of the resulting auxiliaries contribute nothing measurable, but they rarely degrade the performance either: the auxiliary weights are learned as homoscedastic uncertainties, so an auxiliary that fails to reduce target error is down-weighted during training and the model reverts toward single-task behaviour. HIC is the only case where the multi-task model has a lower performance than the two-stream model, but with a difference within the uncertainty (Δρ = 0.03).

## 4.3 XGBoost (XB)

A gradient-boosted regression tree model (XGBoost) served as the baseline. It is a single-head by construction, e.g. it does not use any auxiliary labels. The per-residue structural features were pooled over the valid (masked) positions into a per-channel mean, standard deviation, minimum and maximum, and concatenated with the global physicochemical descriptors to give one row per molecule.

The baseline used the same folds, the same target transform, and the same scoring convention as the neural models. Five models were trained per fold from different random seeds, which differ for XGBoost through row and column subsampling, and their predictions averaged into one prediction per molecule before scoring, as for the neural arms. The number of boosting rounds was selected on the validation fold, as early stopping is for the neural models, so every arm is optimistic in absolute terms to the same degree.

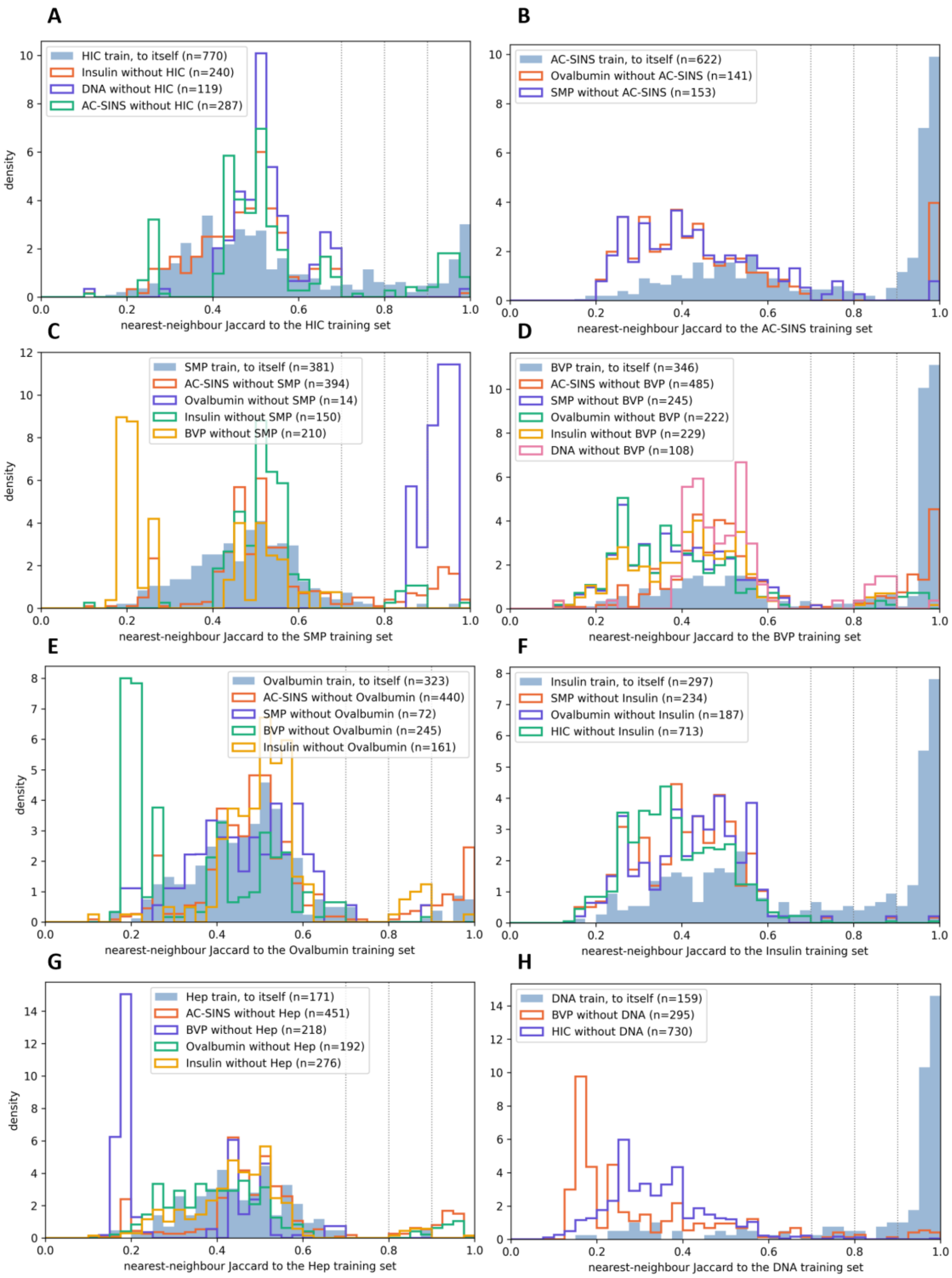

**Fig. 3** Sequence-space structure of each assay's training set, and the auxiliary exposure available from the other assays. One panel per target assay (**A** HIC, **B** AC-SINS, **C** SMP, **D** BVP, **E** Ovalbumin, **F** Insulin, **G** Hep, **H** DNA). Each plot shows the nearest-neighbour k-mer Jaccard similarity of a molecule to the set of molecules carrying target assay (k = 4, VH/VL k-mers pooled; estimated by MinHash-128 hash functions, SE ≈ 0.04). Filled blue: the target's own labelled molecules compared with one another, self-match excluded. Curves- molecules carrying auxiliary assay but **not** the target, compared against the target set. Group sizes are given in each legend; densities are normalised within each curve. Dotted lines mark 0.7, 0.8 and 0.9.

## 4.4 Data splitting for 5-fold cross-validation

As assembled from multiple sources, datasets may contain near-duplicate and closely related sequences, so the commonly used random split can place highly similar antibodies in both the training and evaluation partitions. To avoid this, antibodies were clustered based on pairwise similarity of Fv measured by Jaccard index of k-mer sets[36] (k=4). Two antibodies were linked if *either* their heavy or their light chains reached a similarity threshold of 0.8, and clusters were taken as the connected components of the resulting graph. Heavy and light chains shorter than 90 and 80 residues were assigned an empty k-mer set, have Jaccard 0 with everything, and therefore form singletons.

Sequence similarity for each data set is illustrated in Fig 3. Every assay shows a bimodal structure: a broad mode near 0.45–0.55, corresponding to the diverse sequences, and a second mode at ≥ 0.975 mainly arising from single-mutation series. The latter have relatively large contribution to the AC-SINS, BVP, DNA and Insulin data sets (Fig.3 B, D, H, and F, respectively) and is nearly absent for Heparin (HAC) assay (Fig.3 G). Auxiliary molecules generally lie to the left of the target mass, which indicates, unfortunately, a limited capability of the head target to learn from similar sequences contained in auxiliary labels for available data sets; the clearest exception is AC-SINS for BVP (D), where a subset of AC-SINS-only molecules is near-identical to BVP-labelled molecules.

We performed cross-validation using both random splits, as well as cluster split with grouped partitioning, with every member of a cluster assigned to the same fold. In the latter case, therefore, no pair of sequences split across the training and validation partitions exceeds the 0.8 threshold. Clusters were assigned to folds greedily, largest first, each to the fold currently holding fewest molecules, so that the folds remain comparable in size despite a highly uneven cluster-size distribution.

We distinguish two cross-validation regimes for assessing the multi-task model. In the **auxiliary-imputation** regime a molecule may carry labels in both the target and the auxiliary assays; a validation molecule then remains in the training pool with its auxiliary labels, but its target value is masked. In the **auxiliary-naive** regime the validation molecule is removed from training altogether, including any auxiliary measurements it carries, so nothing about that molecule enters the model. However, auxiliary information can reach it only through other, related sequences. The multi-task results reported under cross-validation (Fig. 5) use the auxiliary-imputation regime.

# 5. Results

## 5.1. Feature importance

To assess feature's contribution across different assays, we measured permutation importance on the Spearman scale for 5 across repeats-fold CV. Each feature was permuted independently with three repeats per fold; the charge-related local and global features were additionally permuted jointly, with five repeats, to measure the importance of the property rather than how it is split across features. Fig. 4 illustrates the result per assay: the mean and the standard deviation across folds with global features are shown in red and local in blue. GPDa3 Specifically, global charge features include pI, total, and also chain charges, while local charge include sum of charge neighbours, the same weighted by exposure, electrostatic dipole moments, electrostatic potential, as well as charged patch area.

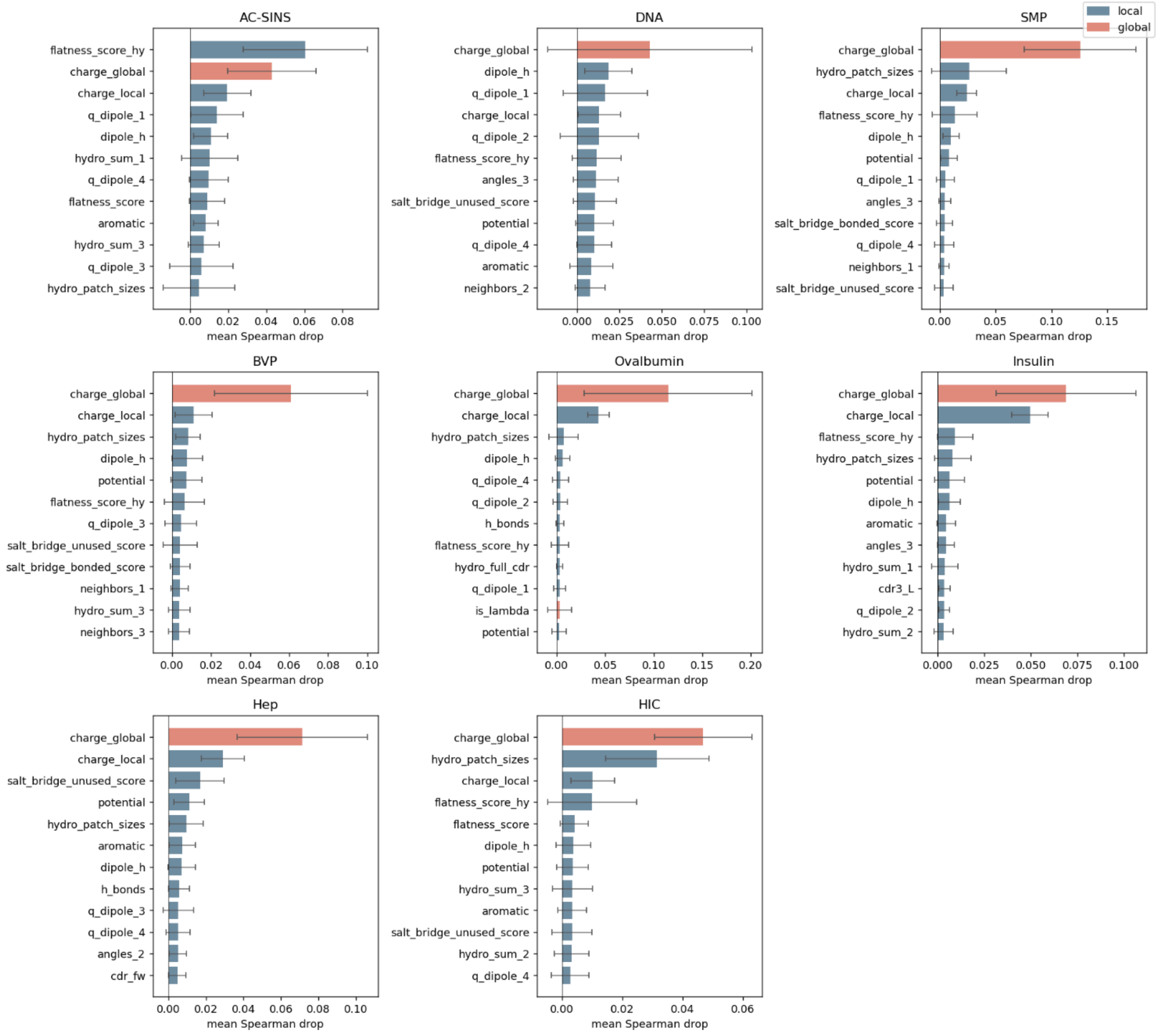

**Fig. 4** Permutation-based feature importance across developability assays. Importance estimation: after training the feature-based model, each feature was randomly shuffled across the antibodies of the validation set and the resulting decrease in Spearman between predicted and measured values was recorded. Each feature was permuted independently, the shuffle was repeated 3 times per fold, and the drop was averaged over repeats. Bars show the mean across the five cluster-based CV folds (5 folds × 3 permutation repeats) and error bars ±1 SD across them. Features are coloured by type: per-residue local descriptors (blue), and global, whole-antibody descriptors (red), which are permuted as one combined block rather than per residue. Only the top 10 features per assay are shown.

Global charge is the largest single contributor for every assay except AC-SINS, where hydrophobic patch flatness ranks first. Local charge, hydrophobic patch size (SMP, HIC) and patch flatness (AC-SINS, Insulin, HIC) follow, and the ordering within that group varies by assay. For the polyspecificity panel charge dominates throughout, indicating that non-specific binding is largely electrostatic. For HIC the hydrophobic terms contribute comparably to charge, as expected for adsorption to a hydrophobic matrix. The AC-SINS exception may reflect that self-association requires a complementary surface patch rather than net charge alone.

These results are consistent with literature. Multiple studies conclude that pI specifically [37],[13] that antibodies with high isoelectric points (pI) have an increased likelihood of high levels of antibody nonspecific interactions, while balance between hydrophobic and charged patches has considerable effect on multiple developability properties Ref.[8]

## 5.2. Benchmark on the GDPa1 dataset

Ginkgo GDPa1[18] benchmark set provides a predefined IgG_isotype_stratified_fold assignment together with published reference scores, so that performance of different models can be compared on identical splits.

We evaluated our single-stream and two-stream models on AC-SINS, HIC and PR_CHO- polyreactivity against soluble membrane proteins from CHO cells (the assay referred to as SMP elsewhere in this work) and compared them with the three reference models reported in the benchmark repository: ablang2_elastic_net, an elastic-net regression on AbLang2 embedding, and moe_baseline, built on MOE descriptors; and CrossAbSense, which couples frozen protein language model encoders to configurable attention decoders[15]. For this comparison our models were trained on the GDPa1 dataset alone, with no other sources included.

Our AbLang2 regressor matches ablang2_elastic_net on HIC (0.46) and exceeds it on AC-SINS (0.60 vs 0.51) and PR_CHO (0.49 vs 0.36). The convolutional feature model exceeds moe_baseline on AC-SINS (0.46 vs 0.42) and PR_CHO (0.45 vs 0.35) but falls below it on HIC (0.53 vs 0.65), the assay whose physics the MOE descriptor set most directly captures.

A consistent pattern is observed: for HIC the feature-based model is the stronger of the two representations (0.65 and 0.53 against 0.46 for either sequence-based model), whereas for self-association and polyreactivity the ordering reverses. This is what the two-stream model is designed to exploit, and it improves on the better of its own two streams for every assay by Δρ = 0.08 on HIC, 0.04 on PR_CHO and 0.01 on AC-SINS. It exceeds CrossAbSense on AC-SINS (0.61 vs 0.48) and PR_CHO (0.53 vs 0.48), and falls slightly below it on HIC (0.61 vs 0.64).

**Table 3** Benchmark on GDPa1 Ginkgo antibody dataset using the predefined IgG_isotype_stratified_fold *: Spearman ρ (5-fold CV average). The best performing model for each assay/model type is shown in bold.

| | Ablang2-based models | | 3D feature-based models | | | |
|---|---|---|---|---|---|---|
| | ablang2+ elastic_net [18] | Ablang2 regressor | MOE_baseline[18] | Features_conv | 2-streams | CrossAbSense [15] |
| HIC | 0.46 | 0.46 | **0.65** | 0.53 | 0.61 | **0.64** |
| AC-SINS (pH7.4) | 0.51 | **0.60** | 0.42 | **0.46** | **0.61** | 0.48 |
| PR_CHO (SMP) | 0.36 | **0.49** | 0.35 | **0.45** | **0.53** | 0.48 |

* https://github.com/ginkgobioworks/abdev-benchmark

## 5.3. Model performance (5-fold split)

Five-fold cross-validation results for five models (XGBoost, each stream alone, two-stream model and the multi-task (MT) model) are shown for each assay in Fig. 5, under random- (A) and cluster-based splitting (B). In both panels the multi-task model is evaluated in the auxiliary-imputation regime: a validation molecule's own auxiliary measurements remain in training and only its target label is withheld. Auxiliary assays are given in Table S3

**Effect of splitting.** Under random splitting the best model exceeded ρ= 0.7 for every assay except AC-SINS (0.64), with the highest value for Insulin (0.87). Cluster splitting separates the assays into two groups by the model performance change: for HIC, Heparin, Ovalbumin and SMP it is essentially unaffected by splitting type- each decreasing by less than Δρ = 0.05. Performance for AC-SINS, BVP, DNA and Insulin decrease substantially (Δρ = 0.12, 0.33, 0.25 and 0.14 respectively) and show markedly wider spread across folds. These four assays are those with a strong association between sequence cluster and data source (NMI > 50 %, Table 1) and a low proportion of singleton sequences (<50%). For them a held-out cluster is usually also a held-out source, so the fold measures transfer to differently calibrated measurements as well as to dissimilar antibodies. These two effects cannot be separated and, thus, not well appropriate for assessment on the model generalizability. The relative ordering of models is largely independent of the splitting scheme.

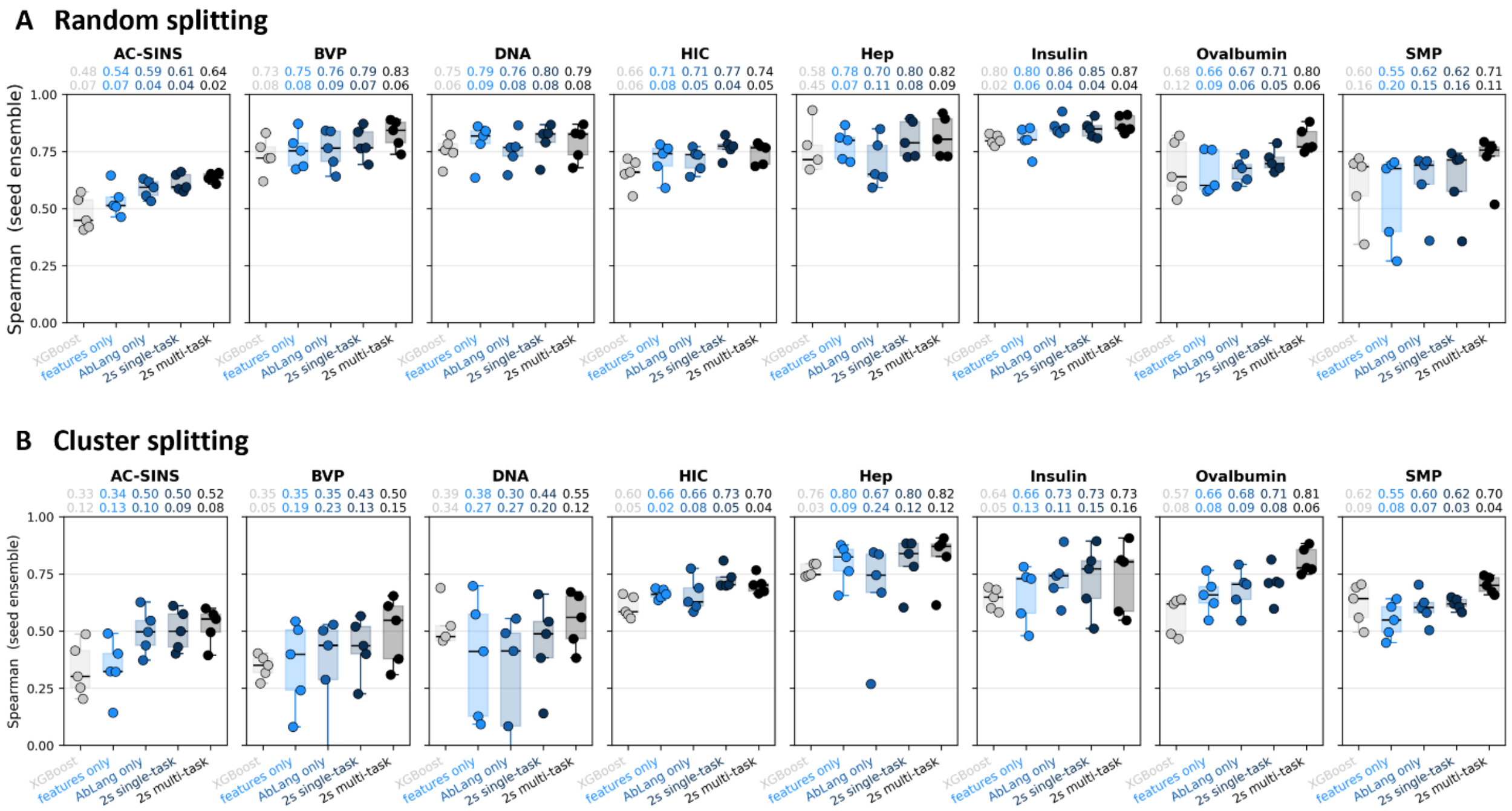

**Fig. 5** Performance of five prediction models across eight developability assays, under 5-fold random (A) and cluster-based (B) splitting. The multi-task model is evaluated in the auxiliary-imputation regime in both panels. The models compared are XGBoost, each stream alone (features only, AbLang only), the two-stream single-task model and the two-stream multi-task model. Each point is one cross-validation fold: predictions from the five random seeds were averaged first, and Spearman correlation computed on that ensemble. Boxes show the median and interquartile range across the five folds. Numbers above each panel give the mean Spearman across folds (upper row) and its standard error (lower row).

**Comparison across models**. One- and two-stream variants improve on the XGBoost baseline for most assays. The AbLang and feature-based streams achieve comparable performance, within fold-to-fold uncertainty, despite being built on an entirely different basis. Combining them is beneficial for most assays, with the largest advantage being for HIC (Δρ = 0.06). Although this improvement is moderate, the two-stream model consistently reduces variability across folds for most assays, indicating improved robustness and better generalisation.

The blend weight is fitted on training loss and therefore reflects which stream fits the training data better rather than which generalises. Fixing it at 0.5 improved AC-SINS, whose training set is dominated by series of near-identical patent-derived variants that the language-model stream fits cheaply, but degraded HIC, where 63 % of training molecules are singletons and the learned weight correctly favours the structural stream.

The multi-task model gives the best performance wherever informative auxiliary assays are available (Sec. 4.2.5). Three conditions on the dataset must hold for information to transfer from auxiliary to target: auxiliary assay available for sequences close to those being predicted, some correlation between target and auxiliary labels on co-measured sequences, and an overlap large enough relative to the target panel for that correlation to carry weight in the loss. Accordingly, the largest and most consistent gains are for the polyreactivity assays, Ovalbumin and SMP (Δρ = +0.09 under random splitting, +0.10 and +0.08 under cluster splitting), which share 309 sequences and correlate strongly with one another (ρ = 0.75). BVP benefits more modestly (+0.04-0.07) despite weaker correlation with its auxiliaries (maximum ρ = 0.48, with SMP), and DNA gains +0.11 under cluster splitting while showing none under random splitting. For AC-SINS, Insulin and Heparin the difference is within fold-to-fold scatter. HIC is the clear case of a target with no suitable auxiliary: the multi-task model falls slightly below the two-stream model under both splitting schemes (-0.03). Its strongest auxiliary, Insulin, correlates at r ≈ 0.5 but only on the 58 sequences measured in both, against 771 HIC measurements. The auxiliary therefore supplies information about under 8 % of the training set, that is too small fraction to change model performance.

**Data accuracy ceiling.** Several assays appear limited by measurement reproducibility rather than by the model. Although we attempted to bring the data onto a common scale, this was not always possible. The lowest inter-source agreement in the present dataset is for AC-SINS (ρ = 0.68), SMP (ρ = 0.72) and Ovalbumin (0.78). The multi-target model performs close to that limit: under cluster splitting it reaches ρ = 0.52 on AC-SINS, 0.70 on SMP and 0.81 on Ovalbumin (Fig. 5). HIC is the opposite case: its measurements agree at 0.98 while the model reaches ρ = 0.73, leaving substantial headroom.

**Classification performance for removing liabilities**. Although rank correlation is modest for several assays, the models separate problematic from acceptable molecules well. Under cluster-based cross-validation AUC lies between 0.80 and 0.97 for every assay except BVP (0.73), and is largely retained across splitting schemes (Fig. 6). These values are adequate for the practical use case of removing liabilities early.

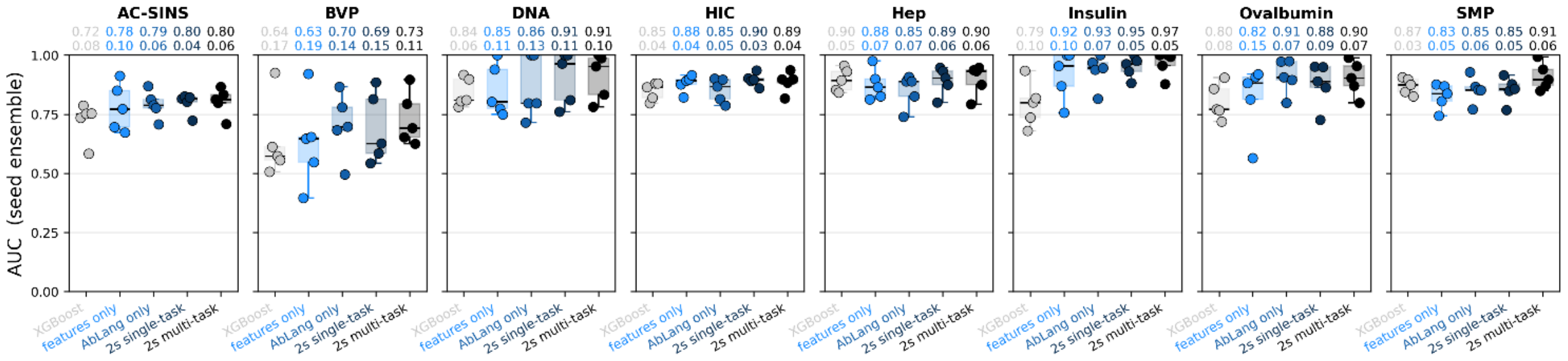

**Fig. 6** Performance comparison of five prediction methods using 5-fold cluster out splitting: the same methods and assays as in Fig.5; point shows AUC (threshold value for each assay used are from Table 1)

**Choice of language model.** Repeating training with ESM2 instead of AbLang2 produced no considerable change in performance; differences for all assays remained within the standard error across folds. This suggests the sequence stream contributes generic sequence information rather than an antibody-specific representation. Practically, the architecture is not tied to a particular pre-trained model and the choice can be made on cost and availability. Note that ESM2 embeddings are standardised on the training fold before use, as the two models differ by an order of magnitude in output scale.

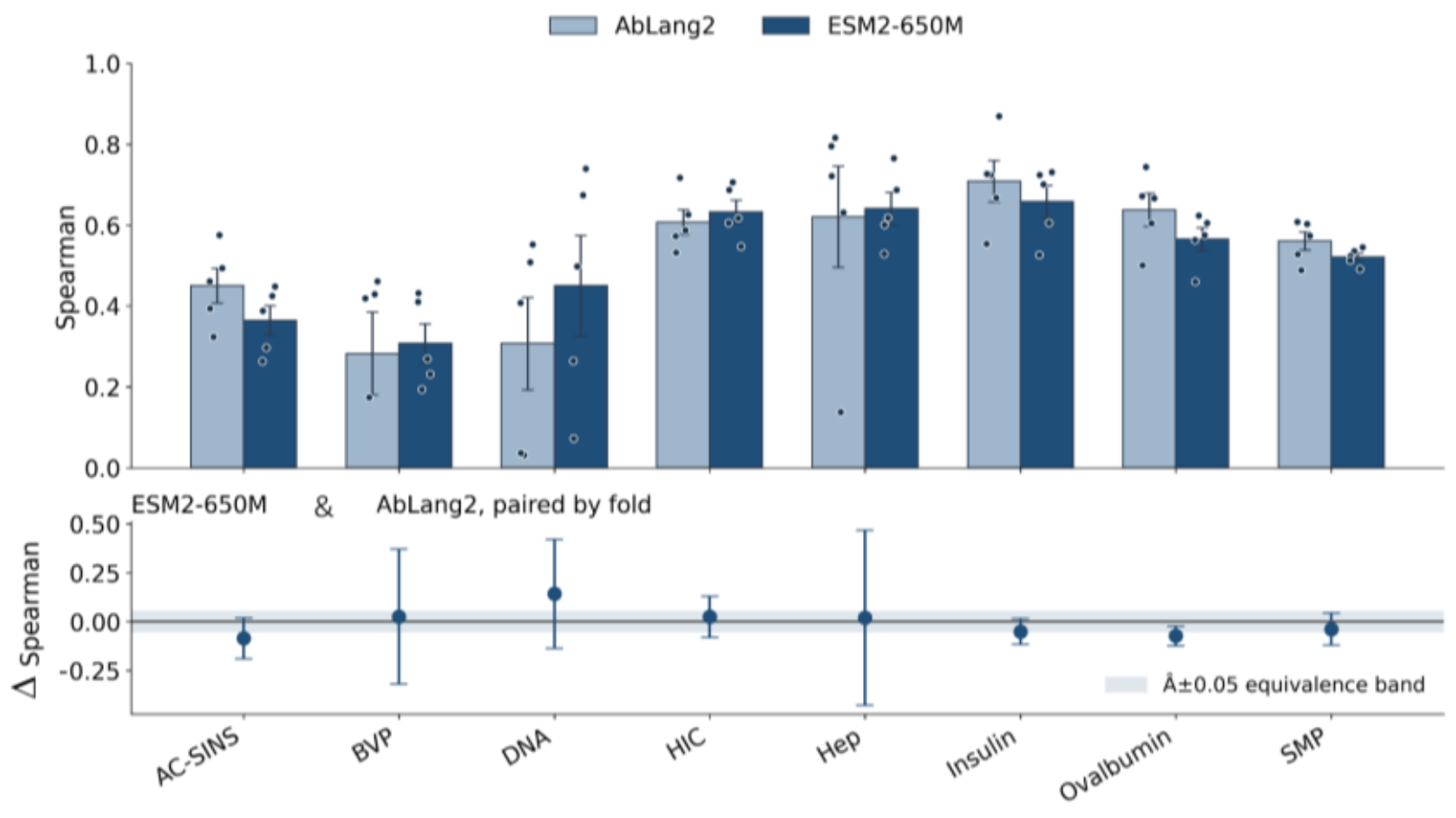

**Fig. 7** Performance comparison of single stream model AbLang versus ESM2 for 8 assays: mean Spearman across folds and its standard error.

## 5.4. Validation on a hold-out GDPa3 set

We first validated the model on the Ginkgo GDPa3 panel[9], which comprises 80 IgGs drawn from the Observed Antibody Space (OAS)[38] and reports HIC, HAC, AC-SINS, PR-CHO and ovalbumin binding. The panel was assembled for maximal diversity and consists entirely of singleton sequences (Table 1). Similarity to the training set is correspondingly low for every assay: the k-mer Jaccard coefficient to the nearest training molecule lies mostly between 0.2 and 0.5 (Fig. 8, right panels), that is well below the 0.8 threshold used to define clusters, so no validation sequence belongs to any training cluster. The contrast with the training sets themselves is clearest for AC-SINS, where a large fraction of training molecules has a near-identical neighbour, while no GDPa3 molecule does. Because there are no sequences in the training set that are close enough to a GDPa3 antibody for an auxiliary measurement, the multi-task route cannot contribute here and the two-stream model is used throughout this section (cross-validation plots for 2-stream models are in Fig. S1).

Performance is highest for HAC and ovalbumin (ρ = 0.69 and 0.67) and lowest for AC-SINS and SMP (0.43 and 0.40), with HIC intermediate (0.65). On AC-SINS and SMP the model is slightly above the best entries of the 2025 Ginkgo Datapoints Antibody Developability Competition (0.36 and 0.34, Table 4), while on HIC it does not reach them (0.63 vs 0.71). No assay exceeds ρ = 0.7, so the predictions are not accurate enough to rank candidates reliably on an unseen panel of this diversity. For the practical task of removing liabilities, however, the classification view is more favourable: AUC = 0.89 for ovalbumin and 0.73 for SMP, against 0.66 for AC-SINS (Table 4).

The largest loss from cross-validation to hold-out is for SMP (Δρ = 0.22). Both CV and hold-out validation plots show the same pattern: the model struggles with correct prediction of the labels close to zero (Fig. 8D, Fig. S1). The GDPa3 panel spans only the lower part of the training range with values reach about 0.45 against 1.0 in training. Therefore, the hold-out asks the model to order molecules within the region where the assay resolves least.

**Table 4** Cross-validated performance of the two-stream (2s) model under cluster splitting (Fig.5) and validation on three external sets: 80 Fv from the Ginkgo GDPa3 panel[9]; cross-format validation on the therapeutic VHH set of Gordon et al. [39] and 160 diverse VHH sequences of OAS[38] from the Ginkgo GDPa5 panel[9]. Values are Spearman ρ and AUC computed with threshold defined in Table1. AUC is given only where the training and validation measurements are on the same scale. Bold marks the higher of our value and the corresponding reference entry.

| | Cluster CV (2-stream) | | Fv- GDPa3[9] | | | VHH of therapeutic antibodies | | VHH- GDPa5[9] | | |
|---|---|---|---|---|---|---|---|---|---|---|
| | ρ | AUC | ρ | ρ* | AUC | ρ | AUC | ρ | ρ** | AUC |
| HIC | 0.73 | 0.90 | 0.65 | **0.71** | - | 0.63*** | - | 0.57 | **0.63** | - |
| AC-SINS (pH7.4) | 0.50 | 0.80 | **0.43** | 0.36 | 0.66 | 0.55*** | 0.88 | 0.38 | **0.65** | 0.75 |
| Ovalbumin | 0.71 | 0.88 | 0.67 | - | 0.89 | - | - | 0.12 | **0.42** | - |
| HAC | 0.80 | 0.89 | 0.69 | - | - | 0.93**** | - | 0.83 | **0.84** | - |
| SMP | 0.62 | 0.85 | **0.40** | 0.34 | 0.73 | - | - | 0.32 | **0.51** | - |

* The best results from [9]

** From Ref. [23] : trained on IgGs data from an internal Ginkgo data set, best model

*** VHH-Gordon et al. [39]

**** GDPa5[9]

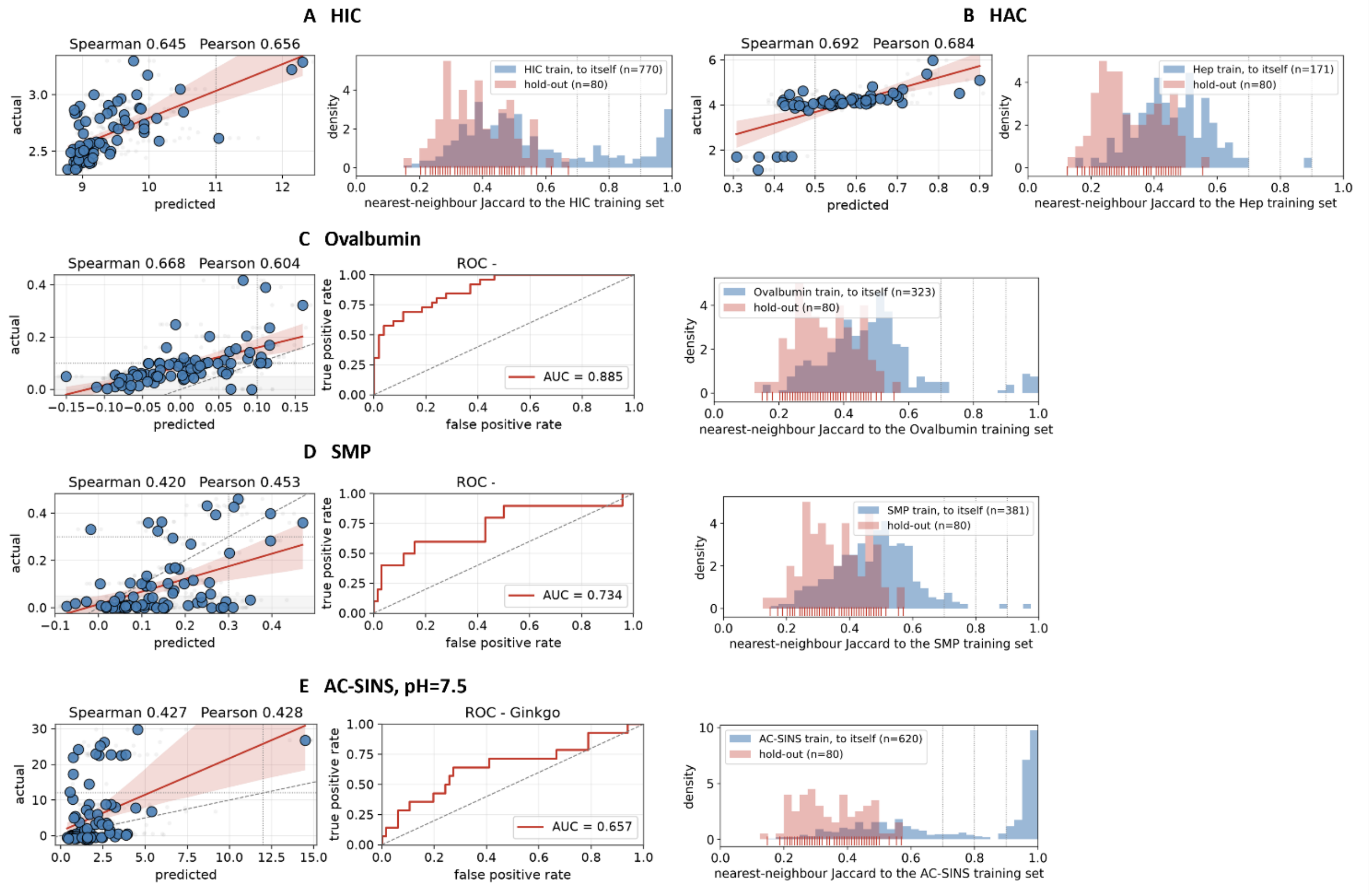

**Fig. 8** **Hold-out validation of the two-stream model on the GDPa3**[9] **dataset.** Five assays: HIC (A), HAC (B), Ovalbumin (C), PR-CHO (SMP, D), and AC-SINS (E). **Left panels (A–E):** Grey points show the predictions of each individual seed (5 seeds), and blue points show ensemble mean across seeds. The red line is the linear best fit with its shaded 95% confidence band; the dashed grey line marks the ideal y = x relationship. **Middle panels** (C-E): ROC curve, **Right panels (A–E):** nearest-neighbour 4-mer Jaccard similarity to the training set. Blue, training molecules to each other (self-excluded); red, hold-out molecules to training. A value of 1.0 indicates an identical sequence in the training set. For HIC, we excluded data points at 4.8 min, as they likely correspond to the instrument's upper detection limit.

## 5.5. Cross-format validation for SMP: an scFv library with binary labels

We next tested SMP model on the data set of Chen et al.[13], comprising over 300,000 diverse naïve human single-chain (scFv) antibodies with polyreactivity labels measured against soluble membrane proteins from CHO cells. The same study also provides quantitative mAb measurements for SMP, insulin and ovalbumin, which are part of our training data; the library sequences used here are distinct from those, as can be seen from the similarity distribution in Fig. 9.

Applying the two-stream SMP model to 3,065 randomly selected sequences, and using threshold of Table 1 we obtain F1 = 0.888, MCC = 0.751 and AUC = 0.947 (Fig. 9), with 89 % of polyreactive and 86 % of non-polyreactive sequences correctly assigned. The model therefore separates polyreactive from clean antibodies in a library setting, in a format it was never trained on, on the basis of 382 quantitative training measurements.

It is noteworthy that, because the binary labels are derived by thresholding, a fraction of the misclassifications may reflect the resolution of the assay near the threshold rather than a failure of the model. Furthermore, scFv and IgG differ in format, and agreement at this level indicates that the predicted signal depends on the variable domains rather than on the constant region or on assay-specific behaviour of the full-length molecule. Conversely, a threshold removes the need to resolve the low-signal labels, which degraded the regression result on GDPa3.

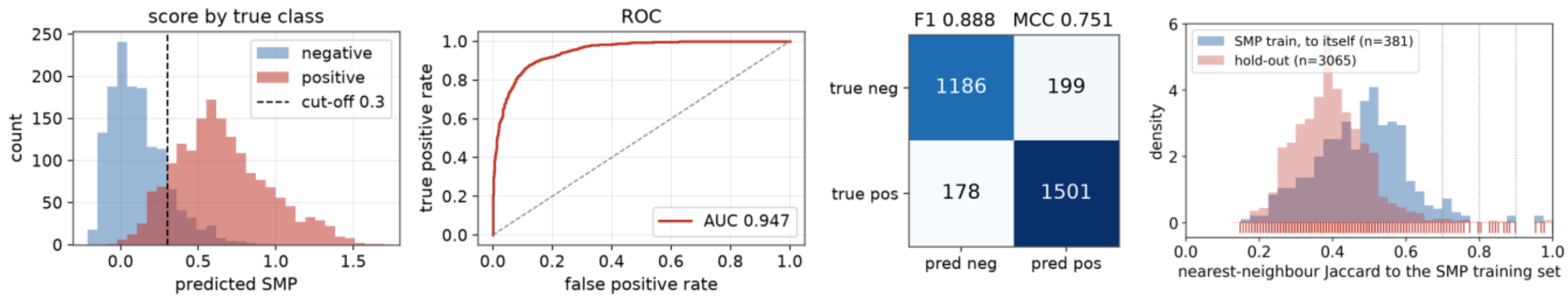

**Fig. 9** Hold-out validation of the two-stream SMP model on the binary dataset of Chen et al.[13] 3 065 sequences were randomly selected from the full dataset (> 300,000 labels), of which 50.8 % are polyreactive. **Left:** distribution of the predicted SMP score by true class, with the cut-off of 0.3 taken from Table 1 (dashed line). **Centre:** ROC curve. **Right of centre**: confusion matrix at that cut-off, corresponding to 89 % of polyreactive and 86 % of non-polyreactive sequences correctly assigned. **Right**: nearest-neighbour 4-mer Jaccard similarity to the training set shown in blue, training molecules to each other (self-excluded); red- hold-out to training; a value of 1.0 would indicate a sequence identical to one in training.

## 5.6. Cross-format inference on VHH sequences

To further test whether an Fv-trained model can be applied to single-domain antibodies, we took HIC as an example and set light-chain-dependent feature channels (two interchain-contact channels and the three light-chain CDR flags) to zero. On randomly selected 24 Fv hold-out sequences, predictions were essentially unchanged (Pearson 0.99 between Fv- and VH predictions; mean absolute shift 1.5% of the prediction spread) and agreement with measurement largely retained (ρ =0.676/0.687, Pearson 0.647/ 0.649 for Fv and VH, respectively). Additionally, setting to zero three channels that sum over six VL CDRs, we estimated the upper bound of the performance reduction (Δρ≈0.03), which indicates that VHH prediction is not dominated by the missing channels in HIC case, and that cross-format transfer preserves ranking.

We then retrained the two-stream models with ESM2 embeddings, since AbLang2-paired is trained on paired VH/VL and has not seen single-domain antibodies, whereas ESM2's training set includes camelid VHHs. These Fv-trained models were applied to the therapeutic VHH measurements of Gordon et al.[39] (HIC, AC-SINS), with HAC values for the same sequences taken from the Ginkgo GDPa5 dataset[23].

Performance is largely retained when moving from the GDPa3 Fv hold-out to this VHH panel. HIC is unchanged (0.63 vs 0.65), while AC-SINS and HAC are predicted better than on GDPa3 (Δρ = +0.12 and +0.24, Table 4). A plausible explanation is that the VHH training corpus contains a substantial number of clinical-stage antibodies, that are closer to molecules in the training set, whereas the GDPa3 panel was assembled from OAS[38] for maximal diversity and lies in unexplored space.

Then the same model was applied to the GDPa5 panel of diverse VHH (they also include re-measured sequences from the Gordon et al.[39] set). Results are comparable to those of Moller and Ritter[23] for HAC (0.83 vs 0.84) and HIC (0.57 vs 0.63), but notably lower for the polyreactivity assays: AC-SINS (0.38 vs 0.65), SMP (0.32 vs 0.51) and Ovalbumin (0.12 vs 0.42). The low accuracy for AC-SINS is expected because of the low consistency between sources in the present data set, while in Ref.[23] model was trained on 559 IgG on the same platform. Similarly, SMP shows consistently problematic discrepancy between training and validation sets already discussed above for the validation on GDPa3 panel. The reason of low performance on Ovalbumin is not clear, as the model showed a good performance on the GDPa3 data set.

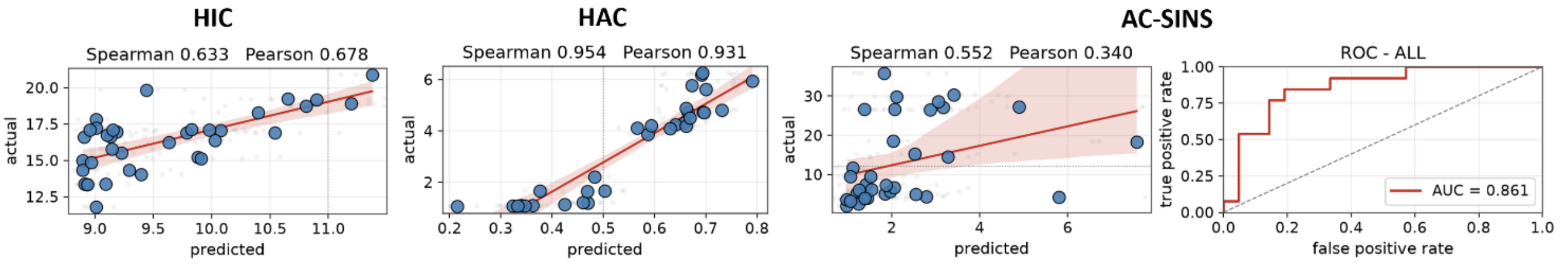

**Fig. 10** Hold-out validation of the two-stream model on therapeutic VHH antibodies: HIC and AC-SINS – measurements from set of Gordon et al. [39], HAC – measurements on the same antibodies from GDPa5[9]

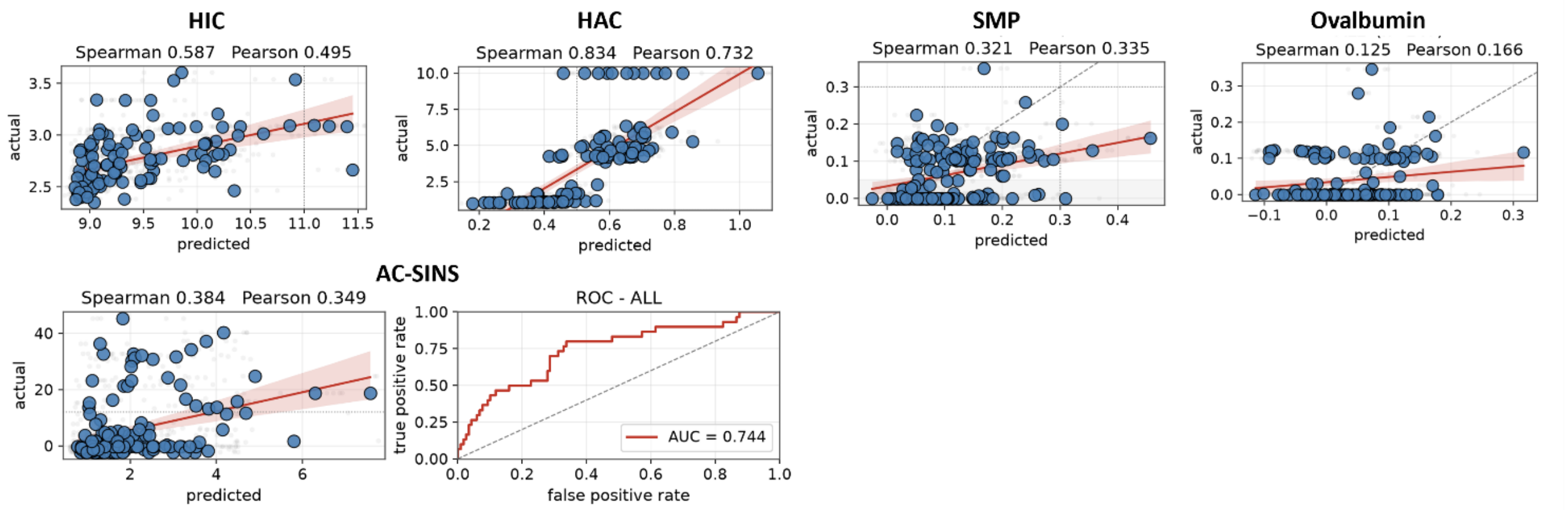

**Fig. 11** Hold-out validation of the two-stream model on GDPa5 VHH panel [9]

# 6. Discussion

## 6.1. Design choices and limiting factors

**The present architecture is designed to work on small datasets**. Three design choices follow from this. First, the protein language model is kept frozen rather than fine-tuned. Second, structural information enters as explicitly designed per-residue descriptors rather than being learned from geometry. Third, the two streams are combined by late fusion of their predictions: the common route is to concatenate a language-model embedding with structure-derived features into a single representation, which allows a shared head to lean on whichever stream is easier to fit, whereas late fusion lets each stream be optimised separately while still being trained jointly.

Larger datasets may permit a finer-grained design. A graph neural network over the residue contact graph can learn spatial relationships directly (Refs. [26], [27]), and the language model can be partially unfrozen or adapted with LoRA rather than used as a fixed encoder (Refs. [27]). The published evidence suggests, however, that the resulting gains are modest. Ahmed et al. [27] train a fine-tuned ESM-2 coupled to a GVP residue-graph encoder with cross-modal disentangled attention on much larger data set (1,499 IgGs and 841 VHHs), while report F1 = 0.57 (IgG) and 0.76 (VHH), against 0.57 and 0.72 for a conventional descriptor-based model trained on the same data. The learned-geometry architecture therefore matches the descriptor model on IgGs and exceeds it by 0.04 on VHHs. The same study applies LoRA and explicit regularisation throughout, "to reduce overfitting on the limited CSI-BLI datasets" that is, at 1,499 molecules overfitting is still the governing concern.

We expect the choices made here to remain appropriate for datasets below roughly a thousand labels per assay. Whether a learned geometric encoder would overtake designed descriptors on these assays given several thousand measurements remains open; the available evidence suggests the margin would be small.

**The multi-task architecture** improves on the two-stream model under 5-fold cross-validation, but the usefulness of an auxiliary assay rests on three conditions: some correlation between the assays on the sequences measured in both ($r \approx 0.5$ is sufficient, as for BVP with SMP); an overlap large enough relative to the target panel for that correlation to carry weight during training; and the presence in the auxiliary data of sequences closely similar to those being predicted. The last condition is not met by panels assembled for maximal diversity, which is why the multi-task model could not be applied to GDPa3 or GDPa5.

**Data.** Though performance should generally improve with larger training sets, it is not considerable for explored data sets. For the descriptor-only model, enlarging the training data from the 246 molecules of GDPa1 to the present 771 (HIC) and 382 (SMP) labels improves ρ by only 0.05-0.08, which indicates that what limits the model is not the number of measurements as such but the coverage of sequence space- molecules far from anything seen in training remain poorly predicted regardless of how many near-duplicates are added.

**Measurement noise** is an equally important limitation. Adding AC-SINS data beyond the Ginkgo GDPa1 set did not improve performance at all. This is partly attributable to inconsistency between sources, which may reflect not only differences in experimental setup but also the use of different Fc regions, IgG1 throughout the additional set, against a mixture of isotypes in the original. Consistent with this, the agreement between sources on molecules measured more than once reaches only 0.68 for AC-SINS, and the persistently low AC-SINS performance reported across models and publications suggests this is a property of the assay rather than of any particular method.

Antibody isotype may act as a scaling factor, though we did not examine this systematically. Because the model uses only the VH and VL sequences, isotype effects would appear as a source-specific offset in the measured values rather than as something the model can resolve. The HIC data suggest the effect is modest: measurements from Shehata et al. [40] and Jain et al. [17] are mutually consistent despite differing in isotype. One route remains untested. Auxiliary assays were shown to help under cross-validation, where a validation molecule's own auxiliary measurements are present (auxiliary-imputation), but we could not asses effect of a related molecule's auxiliary measurement, since that requires a panel in which large set of auxiliary sequence would be close relatives target molecules. No such panel exists in the present corpus, and the external sets were assembled for diversity. Establishing it is left for future work.

# 7. Conclusions

We designed a prediction model for antibody developability assays whose outcomes are largely governed by surface biophysical properties, such as shape, charge distribution and the arrangement of hydrophobic patches. Three questions motivated the design: (i) whether measurements of related (auxiliary) assays, which rest on the same underlying properties, can be used jointly so that one informs the prediction of another; (ii) whether a per-residue representation of those properties can improve prediction on small datasets, opening a route to formats for which no well-established protein language model exists; and (iii) whether data from independent published sources can be pooled to enlarge the training set without the heterogeneity between them cancelling the gain.

The model architecture addresses these by combining, for each assay, a sequence-based stream (a protein language model embedding) and a descriptor-based stream. The latter is a convolutional block over per-residue descriptors computed from a predicted three-dimensional structure. The two streams are combined by fusing their predictions rather than their features, with auxiliary assays entering through shared encoders and separate per-assay heads. We applied the model to eight developability assays: hydrophobic interaction chromatography (HIC), heparin affinity chromatography (HAC), self-association (AC-SINS), and polyspecificity ELISA against BVP, dsDNA, ovalbumin, insulin and soluble membrane proteins. It was then evaluated on four external datasets, two of them in formats absent from the training data: the Ginkgo GDPa3 Fv panel[9], the therapeutic VHH set of Gordon et al. [39], the diverse VHH sequences of GDPa5[23], and an scFv library with binary labels from Chen et al[13]

Validating on the Ginkgo GDPa3 panel we obtain results comparable to, or slightly better than the best competition entries for AC-SINS (0.43 vs 0.36) and SMP (0.40 vs 0.34), while HIC falls below the best entry (0.65 vs 0.71); for the two assays not covered by the competition, Ovalbumin and HAC, we obtain correlations of 0.67 and 0.69.

Evaluation on VHH-Fc demonstrates transferability across formats with model performance largely retained, consistent with the earlier finding of Ref.[23] for another model design. It also shows the limit of generalisation: the 36 therapeutic VHHs of Gordon et al. [39] are predicted more accurately than the 160 diverse VHH sequences drawn from OAS[38], with the largest fall in Spearman for AC-SINS ($\Delta\rho = 0.17$) and smallest for HIC ($\Delta\rho = 0.06$). That HIC transfers best is consistent with its prediction resting predominantly on the descriptor-based stream: generalisation holds where the designed descriptors capture the relevant physics, and depends less on having seen similar sequences. AC-SINS is the opposite case, with large inconsistency across measurements from different groups, which heavily restricts predictive accuracy.

The polyreactivity assays transfer less well, falling well below the models of Moller and Ritter, which were trained on data from the same platform that produced the test labels.[23] We attribute this partly to the heterogeneity of our pooled labels and partly to the difficulty of ranking low-signal measurements, where the assay itself resolves least. It does not, however, prevent the separation of liabilities from acceptable molecules: the same model distinguishes polyreactive from clean sequences in the scFv libraries of Chen et al.[13] with F1 = 0.89 and AUC = 0.95.

Taken together, these results indicate that combining heterogeneous sources and related assays is a viable route to broader and more general developability prediction, and they identify measurement reproducibility and the sequence diversity of the training set, rather than model capacity, as the factors that currently limit it.

## Code availability

Code and released weights are available at https://github.com/DKokh/MADM under the Apache-2.0 licence.

No sequences or experimental measurements are redistributed with the package; all were extracted from publications, publicly available databases or patents, listed with their references in Table S1. The script that rescales measurements to a common baseline, as described in Sec. 3, is included in the package.

## Acknowledgement

The author used Claude (Claude Opus 5, Anthropic) to assist with code generation, debugging, data analysis, and revising the text. All scientific decisions, implementation review, and validation were performed by the author.

# Supplementary Material

## Multi-Assay Antibody Developability Prediction with Shared Structure-Language Representations

Daria B. Kokh

## Content

**Table S1** Data sources (also in the repository, source.xlsx)

**Table S2** Hyperparameters used for 5-fold CV model assessment

**Table S3** Auxiliary assay used for 5-fold CV model assessment

**Fig S1** Training and CV for two-stream model used for validation on GDPa3 data set: left – out of fold per source, right – out of fold – cluster split. Assays: (A) HIC, (B) SMP, (C) HAC, (D) Ovalbumin, (E) AC-SINS

**Table S1 Data sources.** The number of distinct sequences contributed to each assay training after deduplication (in parentheses- the number reported by that source if differs).

| Source | HIC | HAC/ Hep | AC-SINS | BVP | dsDNA | Insulin | Ovalbumin | SMP/ PR-CHO | Ref./ patent |
|---|---|---|---|---|---|---|---|---|---|
| Arsiwala et al. 2025 (GDPa1) | 244 | 44 (93) | 110 (243) | - | - | - | 156 | 196 | [19] |
| Chen et al. 2024 | - | - | - | - | - | 122 | 153 (170) | 148 (205) | [14] |
| Shehata et al. 2019 | 348 | - | - | - | - | - | - | - | [40] |
| Jain et al. 2017 | 4 (135) | - | 137 | 136 | - | - | - | 21 (137) | [18] |
| Patent US20220017622 | - | - | 60 | - | 59 (60) | 59 (60) | - | - | https://patents.google.com/patent/US20220017622A1 |
| Jain et al. 2024 | 43 | 49 | 43 | - | - | - | - | - | 10.1080/19420862.2024.2384104 |
| Patent US20200223907 | - | - | - | 95 (96) | - | - | - | - | https://patents.google.com/patent/US20200223907A1 |
| Patent US20180037644 | - | - | - | 32 (33) | 24 | 23 | - | - | https://patents.google.com/patent/US20180037644A1 |
| Kraft et al. 2020 | - | 78 | - | - | - | - | - | - | 10.1080/19420862.2019.1683432 |
| Patent US20240279346 | 20 | - | - | 20 (23) | 6 | 23 | - | - | https://patents.google.com/patent/US20240279346A1 |
| Patent US20240309070 | - | - | 19 (25) | - | 21 | 21 | - | - | https://patents.google.com/patent/US20240309070A1 |
| Patent US20250019438 | - | - | 28 | 28 | - | - | - | - | https://patents.google.com/patent/US20250019438A1 |
| Patent US20240376225 | - | - | 23 | - | 16 | 16 | - | - | https://patents.google.com/patent/US20240376225A1 |
| Patent US20240059799 | - | - | 54 | - | - | - | - | - | https://patents.google.com/patent/US20240059799A1 |
| Hutchinson et al. 2024 | - | - | 27 | 27 | - | - | - | - | [41] |
| Evers et al. 2024 | 53 | - | - | - | - | - | - | - | [42] |
| Absci 2023 (preprint) | 12 (13) | - | 12 (13) | - | 12 (13) | 12 (13) | - | - | [43] |
| Makowski et al. 2021 | - | - | 30 (31) | - | - | - | 14 (29) | 0 (29) | 10.1080/19420862.2021.1951426 |
| Avery et al. 2018 | - | - | 1 (20) | - | 21 (22) | 21 (23) | - | - | [44] |
| Patent US20240317858 | 17 | - | 17 | - | - | - | - | - | https://patents.google.com/patent/US20240317858A1 |
| Patent US20250019427 | 20 | - | 6 | - | - | - | - | - | https://patents.google.com/patent/US20250019427A1 |
| Huhtinen et al. 2023 | 10 (12) | - | 10 (12) | - | - | - | - | - | [45] |
| Kelly et al. 2015 | - | - | - | - | - | - | - | 17 | [46] |
| Dyson et al. 2020 | - | - | 13 (15) | - | - | - | - | - | [47] |
| Perez et al. 2019 | - | - | 12 | - | - | - | - | - | [48] |
| Dai et al. 2024 | - | - | - | 8 | - | - | - | - | [49] |
| Patent US20240254224 | - | - | 8 | - | - | - | - | - | https://patents.google.com/patent/US20240254224A1 |
| Patent US20240301062 | - | - | 5 | - | - | - | - | - | https://patents.google.com/patent/US20240301062A1 |
| Patent US20250019449 | - | - | 5 | - | - | - | - | - | https://patents.google.com/patent/US20250019449A1 |
| **Total** | 771 (905) | 171 (220) | 620 (784) | 346 (351) | 159 (162) | 297 (301) | 323 (355) | 382 (584) | |

Table S2 Hyperparameters used for 5-fold CV model assessment

Head simple applies to both heads. Where it is True the head has no hidden layer, so Head and Ab head have no effect and are shown as - ; ST/MT is the feature-trunk width for single-task and multi-task runs.

| Assay | Feature trunk | | | Head | Ab trunk | | | |
|---|---|---|---|---|---|---|---|---|
| | Hidden ST/MT | Pool | Head | simple | Ab_model | Ab_pool | Ab hidden | Ab head |
| HIC | 48/48 | attn | 16 | False | conv | attn | 64 | 32 |
| AC-SINS | 32/32 | attn | - | True | small | meanmax | 64 | - |
| BVP | 48/48 | attn | - | True | small | meanmax | 64 | - |
| SMP (PR-CHO) | 48/48 | attn | - | True | small | meanmax | 64 | - |
| Ovalbumin | 16/32 | attn | - | True | small | meanmax | 64 | - |
| Insulin | 16/32 | attn | - | True | small | meanmax | 64 | - |
| dsDNA | 16/32 | attn | - | True | small | meanmax | 64 | - |
| Heparin (HAC) | 16/32 | attn | - | True | small | meanmax | 64 | - |

Table S3 Auxiliary assay used for 5-fold CV model assessment

| Parameters | Auxiliary assays |
|---|---|
| HIC | Insulin, DNA, AC-SINS |
| AC-SINS | Ovalbumin, SMP/PR-CHO, Hep/HAC |
| BVP | Ovalbumin, SMP/PR-CHO, DNA, AC-SINS, Insulin |
| SMP (PR-CHO) | Ovalbumin, BVP, Insulin |
| Ovalbumin | SMP/PR-CHO, Insulin |
| Insulin | Ovalbumin, SMP/PR-CHO, HIC |
| dsDNA | BVP, HIC |
| Heparin (HAC) | AC-SINS, Ovalbumin, BVP, Insulin, HIC |

**Fig S1** CV for 2-stream models used for validation on GDPa3 data set: (**left**)- per source, (**right**)- cluster split.
Assays: (A) HIC, (B) SMP, (C) HAC, (D) Ovalbumin, (E) AC-SINS

A. HIC

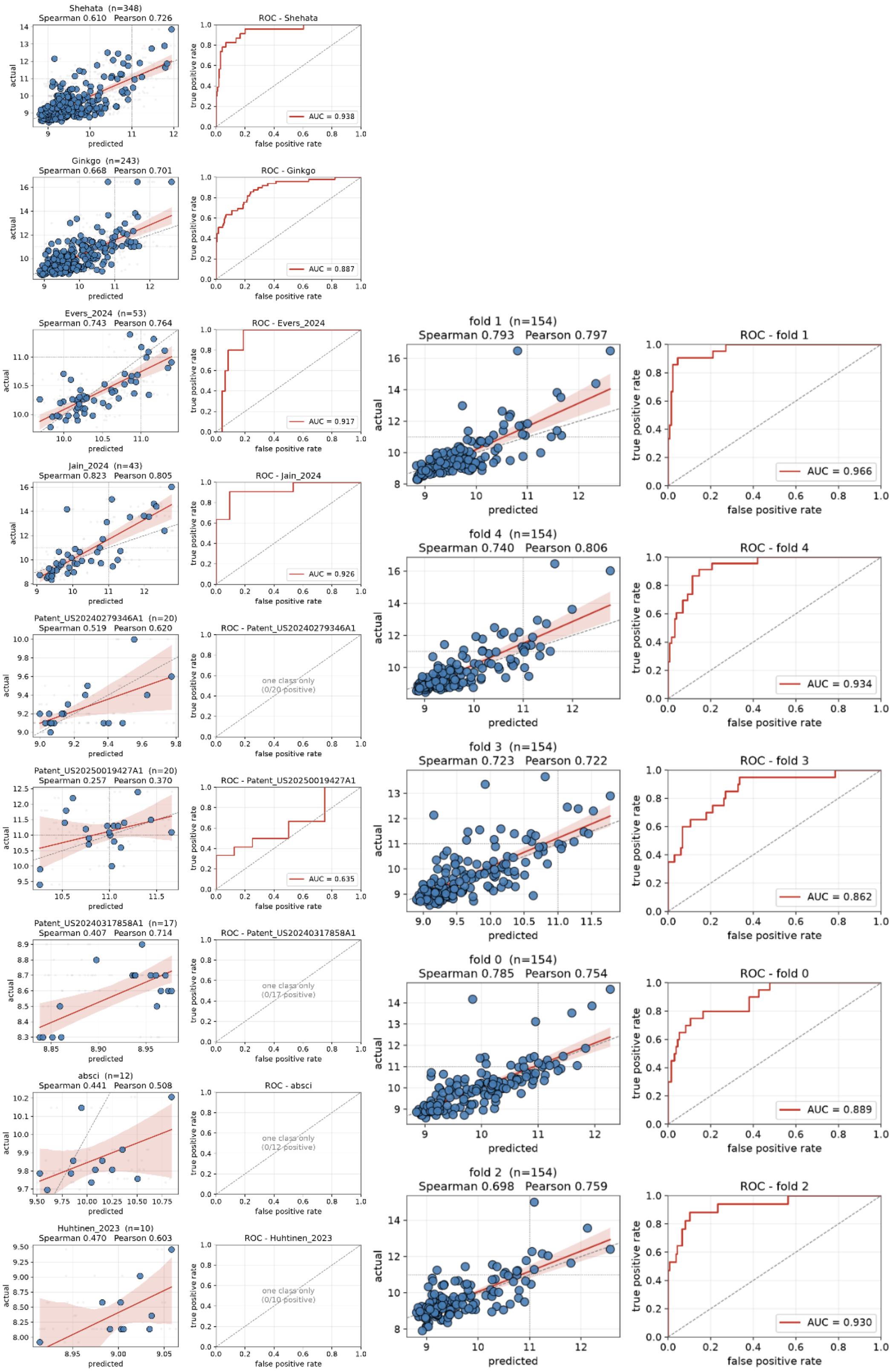

B. SMP

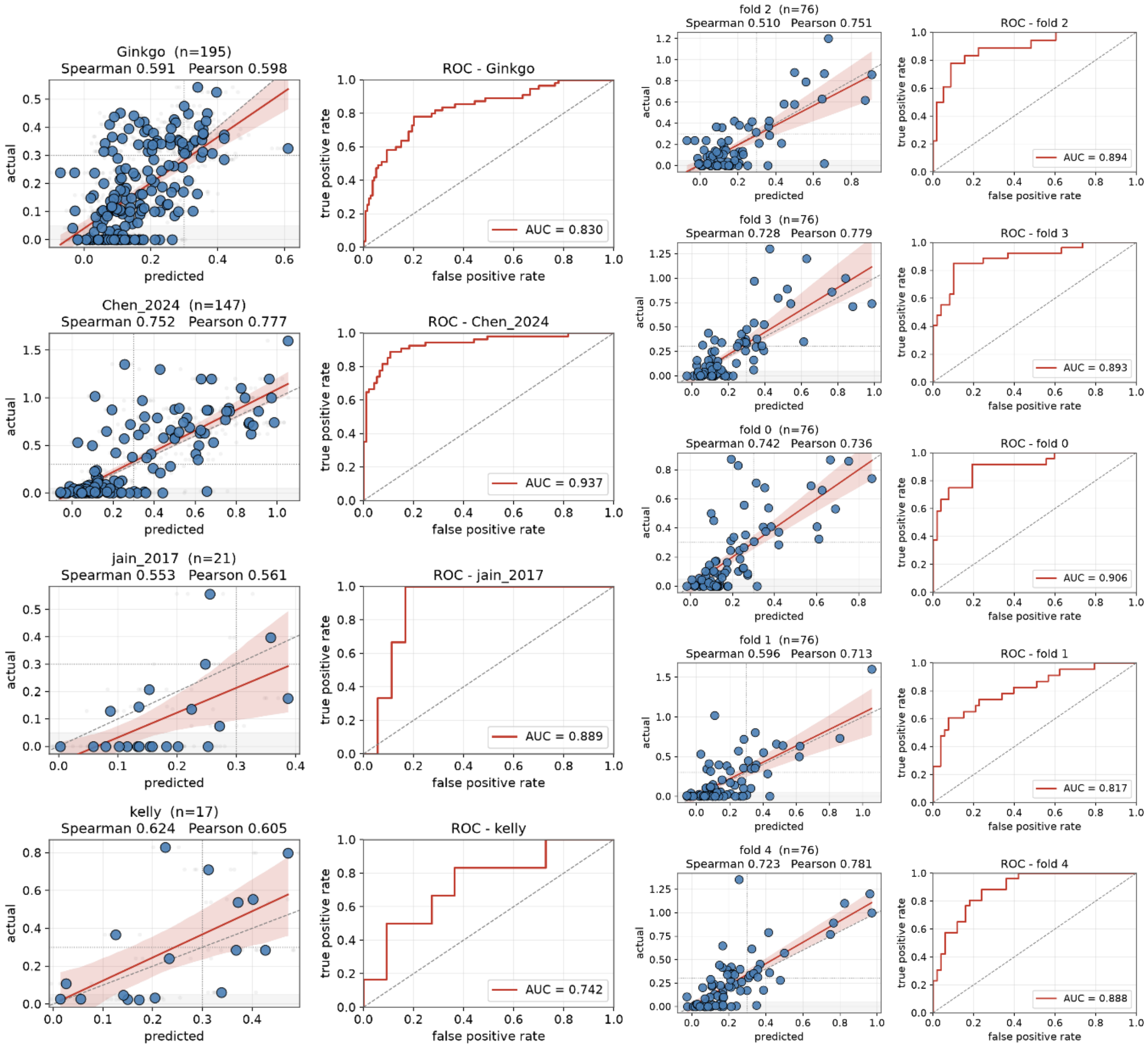
Ginkgo (n=195)
Spearman 0.591 Pearson 0.598
ROC - Ginkgo
AUC = 0.830
Chen_2024 (n=147)
Spearman 0.752 Pearson 0.777
ROC - Chen_2024
AUC = 0.937
jain_2017 (n=21)
Spearman 0.553 Pearson 0.561
ROC - jain_2017
AUC = 0.889
kelly (n=17)
Spearman 0.624 Pearson 0.605
ROC - kelly
AUC = 0.742
fold 2 (n=76)
Spearman 0.510 Pearson 0.751
ROC - fold 2
AUC = 0.894
fold 3 (n=76)
Spearman 0.728 Pearson 0.779
ROC - fold 3
AUC = 0.893
fold 0 (n=76)
Spearman 0.742 Pearson 0.736
ROC - fold 0
AUC = 0.906
fold 1 (n=76)
Spearman 0.596 Pearson 0.713
ROC - fold 1
AUC = 0.817
fold 4 (n=76)
Spearman 0.723 Pearson 0.781
ROC - fold 4
AUC = 0.888
actual
predicted
true positive rate
false positive rate

C. HAC (Heparin binding)

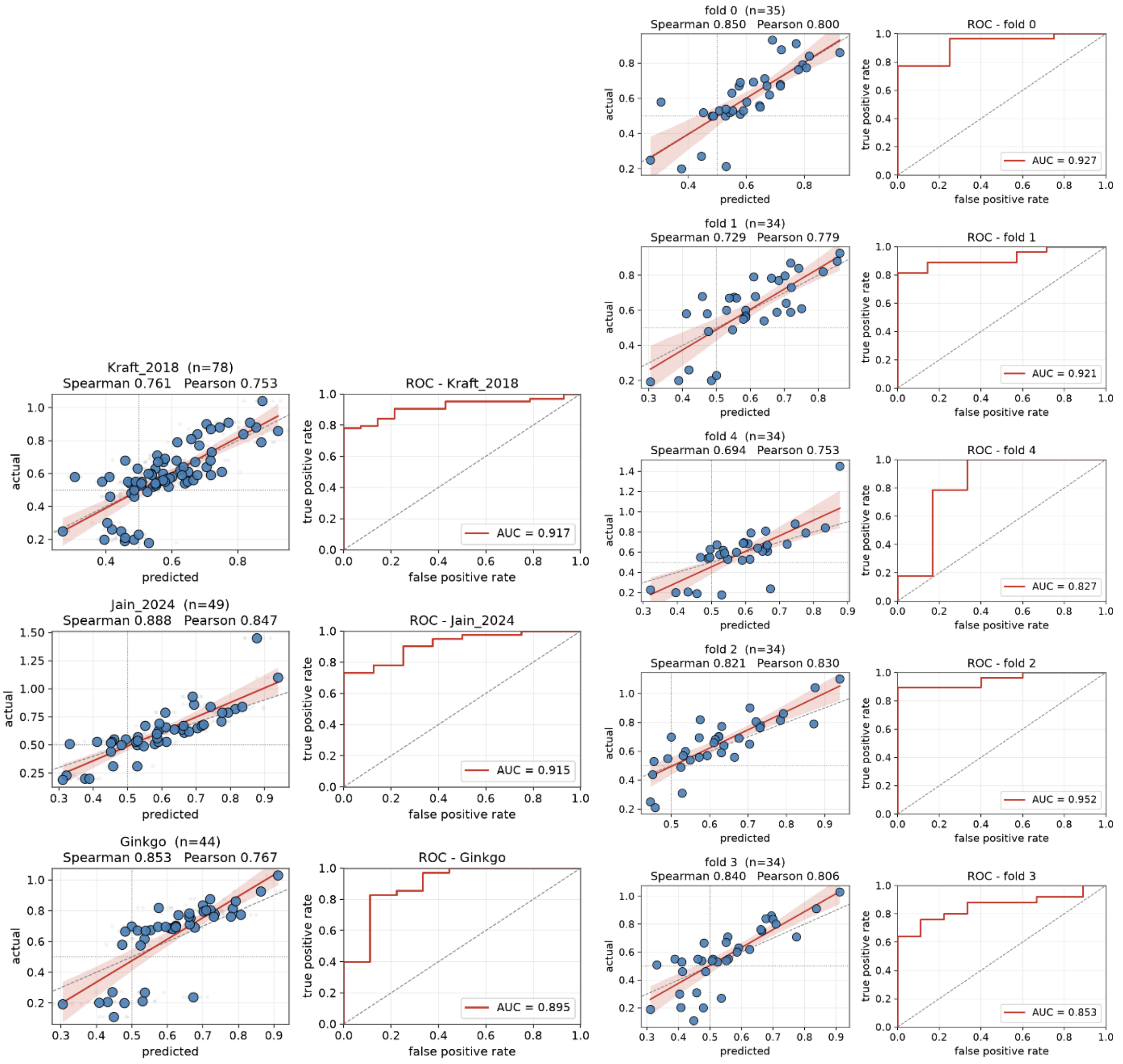

## D. Ovalbumin

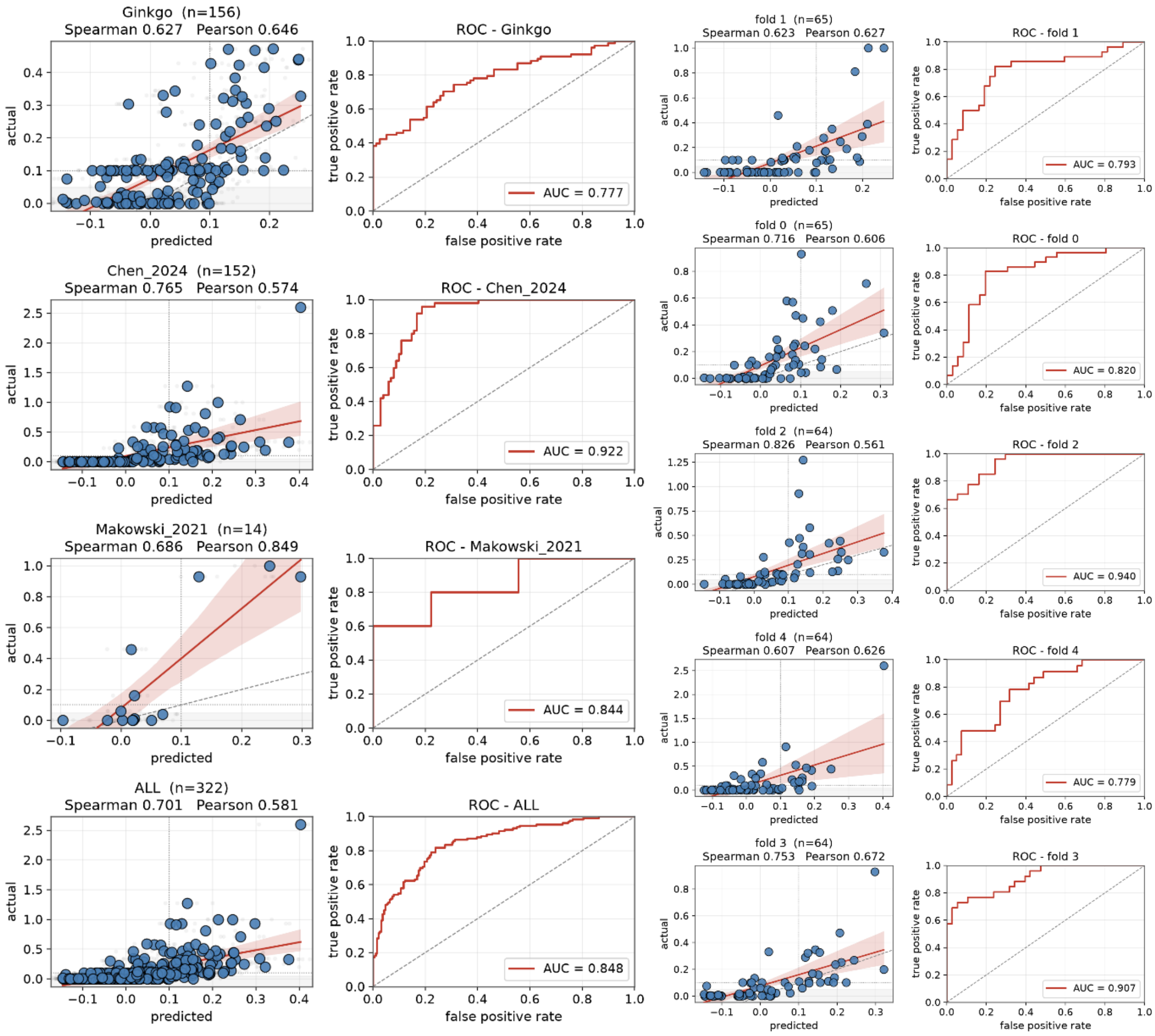

Ginkgo (n=156)
Spearman 0.627 Pearson 0.646
actual
predicted
ROC - Ginkgo
true positive rate
false positive rate
AUC = 0.777
fold 1 (n=65)
Spearman 0.623 Pearson 0.627
ROC - fold 1
AUC = 0.793
Chen_2024 (n=152)
Spearman 0.765 Pearson 0.574
ROC - Chen_2024
AUC = 0.922
fold 0 (n=65)
Spearman 0.716 Pearson 0.606
ROC - fold 0
AUC = 0.820
Makowski_2021 (n=14)
Spearman 0.686 Pearson 0.849
ROC - Makowski_2021
AUC = 0.844
fold 2 (n=64)
Spearman 0.826 Pearson 0.561
ROC - fold 2
AUC = 0.940
fold 4 (n=64)
Spearman 0.607 Pearson 0.626
ROC - fold 4
AUC = 0.779
ALL (n=322)
Spearman 0.701 Pearson 0.581
ROC - ALL
AUC = 0.848
fold 3 (n=64)
Spearman 0.753 Pearson 0.672
ROC - fold 3
AUC = 0.907

# E. AC-SINS

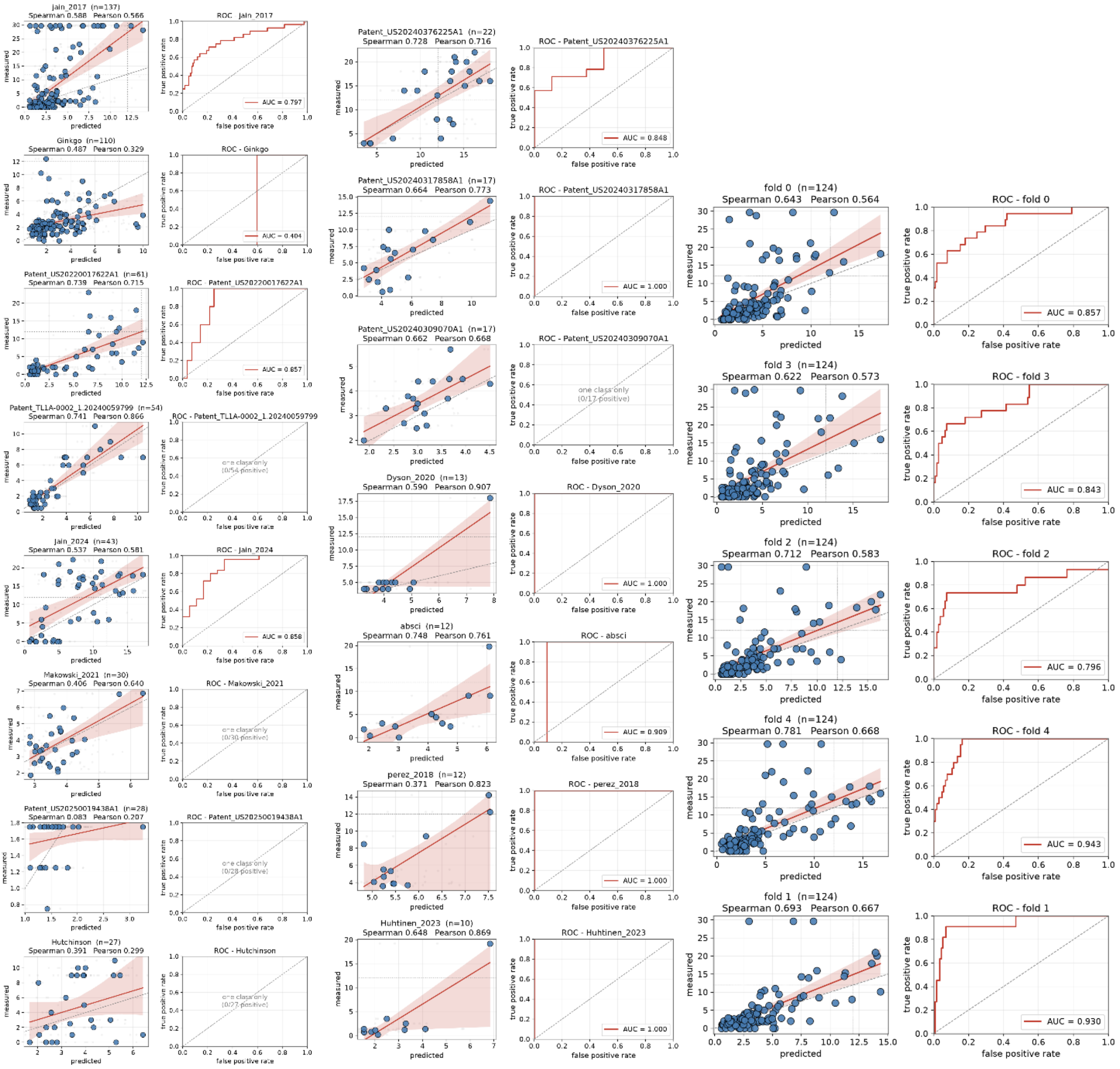
Jain_2017 (n=137)
Spearman 0.588 Pearson 0.566
ROC - Jain_2017
AUC = 0.797
Ginkgo (n=110)
Spearman 0.487 Pearson 0.329
ROC - Ginkgo
AUC = 0.404
Patent_US20220017622A1 (n=61)
Spearman 0.739 Pearson 0.715
ROC - Patent_US20220017622A1
AUC = 0.857
Patent_TL1A-0002_1.20240059799 (n=54)
Spearman 0.741 Pearson 0.866
ROC - Patent_TL1A-0002_1.20240059799
one class only
(0/54 positive)
Jain_2024 (n=43)
Spearman 0.537 Pearson 0.581
ROC - Jain_2024
AUC = 0.858
Makowski_2021 (n=30)
Spearman 0.406 Pearson 0.640
ROC - Makowski_2021
one class only
(0/30 positive)
Patent_US20250019438A1 (n=28)
Spearman 0.083 Pearson 0.207
ROC - Patent_US20250019438A1
one class only
(0/28 positive)
Hutchinson (n=27)
Spearman 0.391 Pearson 0.299
ROC - Hutchinson
one class only
(0/27 positive)
Patent_US20240376225A1 (n=22)
Spearman 0.728 Pearson 0.716
ROC - Patent_US20240376225A1
AUC = 0.848
Patent_US20240317858A1 (n=17)
Spearman 0.664 Pearson 0.773
ROC - Patent_US20240317858A1
AUC = 1.000
Patent_US20240309070A1 (n=17)
Spearman 0.662 Pearson 0.668
ROC - Patent_US20240309070A1
one class only
(0/17 positive)
Dyson_2020 (n=13)
Spearman 0.590 Pearson 0.907
ROC - Dyson_2020
AUC = 1.000
absci (n=12)
Spearman 0.748 Pearson 0.761
ROC - absci
AUC = 0.909
perez_2018 (n=12)
Spearman 0.371 Pearson 0.823
ROC - perez_2018
AUC = 1.000
Huhtinen_2023 (n=10)
Spearman 0.648 Pearson 0.869
ROC - Huhtinen_2023
AUC = 1.000
fold 0 (n=124)
Spearman 0.643 Pearson 0.564
ROC - fold 0
AUC = 0.857
fold 3 (n=124)
Spearman 0.622 Pearson 0.573
ROC - fold 3
AUC = 0.843
fold 2 (n=124)
Spearman 0.712 Pearson 0.583
ROC - fold 2
AUC = 0.796
fold 4 (n=124)
Spearman 0.781 Pearson 0.668
ROC - fold 4
AUC = 0.943
fold 1 (n=124)
Spearman 0.693 Pearson 0.667
ROC - fold 1
AUC = 0.930
measured
predicted
true positive rate
false positive rate